\documentclass[graybox]{svmult}

\usepackage{helvet}         
\usepackage{courier}        
\usepackage{type1cm}        
\usepackage{makeidx}         
\usepackage{graphicx}        
\usepackage{multicol}        
\usepackage[bottom]{footmisc}

\usepackage{amsmath}
\usepackage{amssymb}
\usepackage{braket}
\usepackage{slashed}
\usepackage{bbm}
\usepackage{ifthen}
\usepackage{tikz}
\usetikzlibrary{matrix,arrows,shapes}

\DeclareMathOperator{\Tr}{Tr}

\newcounter{x}
\newcounter{y}
\newcounter{z}
\newcommand\xaxis{210}
\newcommand\yaxis{-30}
\newcommand\zaxis{90}
\newcommand\topside[3]{
  \fill[fill=orange!60!yellow, draw=black,shift={(\xaxis:#1)},shift={(\yaxis:#2)},
  shift={(\zaxis:#3)}] (0,0) -- (30:1) -- (0,1) --(150:1)--(0,0);
}
\newcommand\leftside[3]{
  \fill[fill=orange!70!yellow, draw=black,shift={(\xaxis:#1)},shift={(\yaxis:#2)},
  shift={(\zaxis:#3)}] (0,0) -- (0,-1) -- (210:1) --(150:1)--(0,0);
}
\newcommand\rightside[3]{
  \fill[fill=orange!50!yellow, draw=black,shift={(\xaxis:#1)},shift={(\yaxis:#2)},
  shift={(\zaxis:#3)}] (0,0) -- (30:1) -- (-30:1) --(0,-1)--(0,0);
}
\newcommand\cube[3]{
  \topside{#1}{#2}{#3} \leftside{#1}{#2}{#3} \rightside{#1}{#2}{#3}
}
\newcommand\planepartition[1]{
\begin{tikzpicture}[scale=.4];
\draw[thick,->] (0,0) -- (\xaxis:7.5) node[below] {$x_1$};
\draw[thick,->] (0,0) -- (\yaxis:7.5) node[below] {$x_2$};
\draw[thick,->] (0,0) -- (\zaxis:7.5) node[right] {$x_3$};
 \setcounter{x}{-1}
  \foreach \a in {#1} {
    \addtocounter{x}{1}
    \setcounter{y}{-1}
    \foreach \b in \a {
      \addtocounter{y}{1}
      \setcounter{z}{-1}
      \foreach \c in {0,...,\b} {
        \addtocounter{z}{1}
      \ifthenelse{\c=0}{\setcounter{z}{-1},\addtocounter{y}{0}}{
        \cube{\value{x}}{\value{y}}{\value{z}}}
      }
    }
  }
\end{tikzpicture}  
}
\newcommand\planepartitionsm[1]{
\begin{tikzpicture}[scale=.2];
\draw[thick,->] (0,0) -- (\xaxis:7.5) node[below] {$x_1$};
\draw[thick,->] (0,0) -- (\yaxis:7.5) node[below] {$x_2$};
\draw[thick,->] (0,0) -- (\zaxis:7.5) node[right] {$x_3$};
 \setcounter{x}{-1}
  \foreach \a in {#1} {
    \addtocounter{x}{1}
    \setcounter{y}{-1}
    \foreach \b in \a {
      \addtocounter{y}{1}
      \setcounter{z}{-1}
      \foreach \c in {0,...,\b} {
        \addtocounter{z}{1}
      \ifthenelse{\c=0}{\setcounter{z}{-1},\addtocounter{y}{0}}{
        \cube{\value{x}}{\value{y}}{\value{z}}}
      }
    }
  }
\end{tikzpicture}  
}

\begin{document}

\title*{$\mathcal{W}$-algebras and integrability}
\author{Tom\'{a}\v{s} Proch\'{a}zka}
\institute{Tom\'{a}\v{s} Proch\'{a}zka \at Institute of Physics of the Czech Academy of Sciences, \\ Na Slovance 1999/2, 182 00 Prague 8, Czech Republic \\ \email{prochazkat@fzu.cz}
}
\maketitle

\abstract{This is a short non-technical review focusing on the $\mathcal{W}_N$ family of $\mathcal{W}$-algebras and on their relation to quantum integrability. It is a summary of recently given seminars and workshop contributions.}

\section{Motivation}

$\mathcal{W}$-algebras are higher spin extensions of symmetry algebras of the two-dimensional conformal field theory. Despite the fact that they can be difficult to describe explicitly, their definition is a very natural one and this is one of the reasons why they appear in many contexts in mathematical physics. Let us list few of these to illustrate the usefulness of these algebraic structures:
\begin{itemize}
\item integrable hierarchies of partial differential equations (Korteweg-de Vries hierarchy, Kadomtsev-Petviashvili hierarchy) \cite{Dickey:1991xa}
\item ``old'' matrix models \cite{Itoyama:1991hz}
\item holographic dual description of 3d higher spin theories \cite{Campoleoni:2011hg,Gaberdiel:2010pz,Gaberdiel:2012ku}
\item 4d Nekrasov instanton partition functions and Alday-Gaiotto-Tachikawa correspondence \cite{Alday:2009aq,LeFloch:2020uop,Nekrasov:2002qd,Wyllard:2009hg}
\item BPS subsectors of 4d superconformal field theories \cite{Beem:2013sza,Gaiotto:2017euk}
\item topological strings \cite{Aganagic:2003db,Aganagic:2003qj}
\item quantum Hall effect \cite{Cappelli:1992yv}
\item geometric representation theory -- equivariant cohomology of various moduli spaces \cite{Schiffmann:2012tbu} and combinatorics of plane partitions or lozenge tilings
\end{itemize}

One of the aims of this short note is to summarize some of the recent progress in understanding of these algebraic structures -- focusing on the connections to quantum integrability. There is an excellent review of the status of $\mathcal{W}$-algebras from the beginning of 90s by Bouwknegt and Schoutens \cite{Bouwknegt:1992wg}, but the impotant work of Nekrasov \cite{Nekrasov:2002qd} as well as Alday, Gaiotto and Tachikawa \cite{Alday:2009aq} lead to renewed interest in these structures and significant progress in recent years. In the following we will review some aspects of these developments. The focus is on the interplay between 2d conformal field theory point of view and algebraic structures that are natural from the perspective of quantum integrability. For other related recent developments that are not reviewed here see \cite{Bourgine:2019phm,Coman:2020qgf,Gaiotto:2019wcc,Galakhov:2021xum,Gukov:2022gei,Mironov:2020pcd,Nieri:2019mdl,Rapcak:2021hdh} and the references contained therein.

\section{$\mathcal{W}$-algebras}
\label{secwalgebras}
In this section we will introduce the $\mathcal{W}_N$ family of the vertex operator algebras as well as algebras $\mathcal{W}_\infty$ and $\mathcal{W}_{1+\infty}$ which are the main subjects of this note.

\subsection{Zamolodchikov's $\mathcal{W}_3$}
The oldest genuine example of $\mathcal{W}$-algebra is the algebra $\mathcal{W}_3$ constructed by Zamolodchikov \cite{Zamolodchikov:1985wn} shortly after \cite{Belavin:1984vu}. The starting point was the Virasoro algebra which in the language of operator product expansions is specified by the operator product expansion
\begin{equation}
\label{virasoroope}
T(z) T(w) \sim \frac{c/2}{(z-w)^4} + \frac{2T(w)}{(z-w)^2} + \frac{\partial T(w)}{z-w} + reg.
\end{equation}
Here $c$ is the central charge. Zamolodchikov considered extending the Virasoro algebra by an additional local field $W(z)$ of dimension $3$ which in addition he assumed to be primary with respect to the Virasoro algebra,
\begin{equation}
\label{twope}
T(z) W(w) \sim \frac{3W(w)}{(z-w)^2} + \frac{\partial W(w)}{z-w} + reg.
\end{equation}
In order to close the algebra, it is necessary to find the operator product expansion of $W(z)$ with itself. The form of the OPE is strongly restricted by associativity. The associativity conditions can be phrased in various ways (such as associativity of the algebra of Fourier modes of $T$ and $W$ or as crossing symmetry of the correlation functions etc., see \cite{Bouwknegt:1992wg}), but in the end they all lead to the result
\begin{align}
\label{wwope}
\nonumber
W(z) W(w) & \sim \frac{c/3}{(z-w)^6} + \frac{2T(w)}{(z-w)^4} + \frac{\partial T(w)}{(z-w)^3} \\
& + \frac{1}{(z-w)^2} \left(\frac{32}{5c+22} \Lambda(w) + \frac{3}{10} \partial^2 T(w) \right) \\
\nonumber
& + \frac{1}{z-w} \left( \frac{16}{5c+22} \partial \Lambda(w) + \frac{1}{15} \partial^3 T(w) \right) + reg.
\end{align}
where we introduced a quasiprimary \emph{composite} field of spin $4$
\begin{equation}
\Lambda(z) = (TT)(z) - \frac{3}{10} \partial^2 T(z).
\end{equation}
Note that apart for the overall normalization of the field $W(z)$ which was not specified by \eqref{twope}, all the coefficients in \eqref{wwope} are uniquely fixed by imposing the associativity conditions. We therefore arrive at one-parametric family of $\mathcal{W}_3$ algebras, the parameter $c$ being the central charge just like in the case of Virasoro algebra.

\subsubsection{Non-linearities and $\mathcal{W}$-algebras} The expression \eqref{wwope} is clearly more complicated than the stress-energy tensor OPE and in particular is non-linear due to appearance of dimension $4$ field $\Lambda(z)$ which is a quadratic composite of the generating field $T(z)$. The appearance of non-linearities should not be a surprise: since the fields on the left-hand side of \eqref{wwope} have dimension $3+3=6$, the right-hand side has most singular pole of sixth order and furthermore the local operators appearing in the singular part of the OPE are of dimension up to $5$. For this reason the composite operators such as $(TT)(z)$ of dimension $4$ and higher can and do appear in the OPE. In order words, the fact that affine Lie algebras or Virasoro algebra can be considered simply as Lie algebras instead of more generally as $\mathcal{W}$-algebras, is a simplification that happens because these algebras have generating fields of low dimensions $1$ or $2$. As soon as we have a field of dimension $3$ or higher, the non-linearities cannot generically be avoided and one has to consider the more general notion of $\mathcal{W}$-algebra as a non-linear generalization of a Lie algebra.

\subsection{$\mathcal{W}_N$ family of algebras and their construction}
The Virasoro algebra and $\mathcal{W}_3$ algebra are first two members of a very interesting family of algebras usually denoted by $\mathcal{W}_N$ and associated to $\mathfrak{sl}(N)$ family of simple Lie algebras. $\mathcal{W}_N$ can be defined by extending the Virasoro algebra by primary fields of spin $3, 4, \ldots, N$. For lower values of rank $N$ one can repeat Zamolodchikov's bootstrap analysis, writing the most general ansatz for OPEs of generating fields and imposing the associativity conditions \cite{Prochazka:2014gqa}. But even for moderately low values of rank $N$ these calculations get very complicated due to the fact that the number of fields appearing in the singular part of OPE grows exponentially. In all the situations where one can do such calculation explicitly, one finds that for every integer $N \geq 2$ there exists a one-parameter family of algebras, the parameter being the central charge $c$\footnote{Here we are making two hidden assumptions: first, we are looking for algebras which satisfy the associativity conditions for generic values of the central charge and second we are assuming that the generating fields $W_j(z)$ do not in general satisfy any relations. If we allow for such relations, one finds other branches of solutions of associativity solutions such as the $Y_{012}$ family \cite{Blumenhagen:1990jv,Prochazka:2018tlo} which is gives another family of solutions of the bootstrap with generating fields of spin $2, 3, \ldots, 6$.}. Fortunately, there are alternative constructions of algebras of $\mathcal{W}_N$ family:
\begin{itemize}
\item Perhaps simplest construction is the GKO coset construction \cite{Goddard:1984vk} of so-called Casimir algebra \cite{Bouwknegt:1992wg}. We consider the affine Lie algebra $\widehat{\mathfrak{su}}(N)_1$ (here $1$ denotes the level) and restrict to fields that commute with the global $\mathfrak{su}(N)$ subalgebra. This commutant is usually denoted by
\begin{equation}
\frac{\widehat{\mathfrak{su}}(N)_1}{\mathfrak{su}(N)}.
\end{equation}
It is closed under OPEs and in fact gives $\mathcal{W}_N$ algebra of a specific central charge, $c = N-1$.
\item The coset construction can be slightly generalized. Studying a sum of two affine Lie algebras $\widehat{\mathfrak{su}}(N)_k$ and $\widehat{\mathfrak{su}}(N)_1$\footnote{Note that if we replaced $\widehat{\mathfrak{su}}(N)_1$ by $\widehat{\mathfrak{su}}(N)_\ell$, we would also get a $\mathcal{W}$-algebra, but larger than $\mathcal{W}_N$ \cite{Eberhardt:2020zgt}, with multiplicities of generating fields being more than $1$.} and restricting to the commutant of the diagonal subalgebra $\widehat{\mathfrak{su}}(N)_{k+1}$, we find the coset
\begin{equation}
\frac{\widehat{\mathfrak{su}}(N)_k \times \widehat{\mathfrak{su}}(N)_1}{\widehat{\mathfrak{su}}(N)_{k+1}}
\end{equation}
which realizes the $\mathcal{W}_N$ algebra of central charge
\begin{equation}
c = \frac{k(N-1)(2N+k+1)}{(N+k)(N+k+1)}.
\end{equation}
As $k$ runs over integrable levels, $k=1,2,\ldots$, the corresponding central charges run over all unitary minimal models of $\mathcal{W}_N$ algebras. For example, for $N=2$ and $k=1$ we get the Ising model with $c = \frac{1}{2}$.
\item Another general way of producing $\mathcal{W}_N$ algebra starting with affine Lie algebra is to consider the quantum Hamiltonian reduction (so-called Drinfeld-Sokolov reduction). Here one uses the BRST procedure to impose a constraint on affine Lie algebra. The OPEs of the fields in cohomoloy again close on a $\mathcal{W}$-algebra \cite{Bouwknegt:1992wg,Feigin:1990pn}. Applying this procedure to $\widehat{\mathfrak{su}}(N)_k$ leads to $\mathcal{W}_N$ algebra of central charge
\begin{equation}
c = -\frac{(N-1)(N^2+Nk+k)(N^2+Nk-N-1)}{N+k}.
\end{equation}
We again find that by suitably choosing the level $k$ we can construct $\mathcal{W}_N$ algebra of any central charge. Note given a central charge of $\mathcal{W}_N$ algebra, the level $k$ entering the GKO coset construction and the level entering the DS reduction do not have the same value.
\item Perhaps the most direct way of constructing $\mathcal{W}_N$ algebras is by representing their generating fields in terms of free bosons. To do this, one starts with $N$ commuting free bosons $J_j(z)$ with OPE
\begin{equation}
J_j(z) J_k(w) \sim \frac{\delta_{jk}}{(z-w)^2} + reg.
\end{equation}
and defines the composite fields $U_j(z)$ by Miura transformation
\begin{equation}
\label{miura}
(\alpha_0 \partial_z + J_1(z)) (\alpha_0 \partial_z + J_2(z)) \cdots (\alpha_0 \partial_z + J_N(z)) = \sum_{j=0}^N U_j(z) (\alpha_0 \partial_z)^{N-j}
\end{equation}
It turns out that the fields $U_j(z)$ close under OPE to $\widehat{\mathfrak{u}}(1) \times \mathcal{W}_N$ algebra and are related by a triangular field redefinition to the primary generators $W_j(z)$ discussed above \cite{Bouwknegt:1992wg,Fateev:1987zh}.

The Miura transformation can be thought of as a double quantization of a factorization of a polynomial into a product of simple factors. The free bosons $J_j(z)$ play the role of eigenvalues while the composite fields $U_j(z)$ replace the elementary symmetric polynomials of the eigenvalues. The first quantization corresponds to replacement of the indeterminate by a derivative operator $\partial$ and eigenvalues by functions (classical fields). This way one obtains a free field representation of a classical $\mathcal{W}$-algebra. The second quantization reflects the fact that we do not consider $J_j(z)$ as classical fields but rather as local quantum fields. This picture allows us to view $\mathcal{W}_N$ algebra as quantization of the algebra of $N$-th order differential operators. The Miura transformation is also very imporant element connecting to the dual integrable point of view as we will review later.

As a final comment, note that although the fields $U_j(z)$ are neither primary nor quasiprimary, their OPEs have special feature that the singular part of the OPE involves only their \emph{quadratic} composites. This is a huge simplification in the bootstrap procedure and actually leads to conjectural closed-form formula for all the OPEs in terms of bilocal quantities \cite{Prochazka:2014gqa}.
\end{itemize}

\subsection{$\mathcal{W}_\infty$ and interpolating algebras}
It turns out to be very important not to study the algebras of $\mathcal{W}_N$ family individually, but instead to consider them as members of a two-parametric family parametrized by the rank $N$ and the central charge $c$. The problem is that there is in general no embedding $\mathcal{W}_N \subset \mathcal{W}_{N+1}$. On the other hand, there actually exists a two-parameter family of algebras $\mathcal{W}_\infty[\lambda,c]$ interpolating between all algebras of $\mathcal{W}_N$ family. Its construction can be illustrated by the analogous situation in the context of $\mathfrak{sl}(N)$ family of simple Lie algebras which is well known from the literature on higher spin theories \cite{Campoleoni:2011hg,Gaberdiel:2011wb}.

\subsubsection{Higher spin algebra and fuzzy spheres}

Let us consider the associative algebra of $N \times N$ matrices and let us embed the Lie algebra of $\mathfrak{so}(3)$ as $N \times N$ matrices in $N$-dimensional irreducible representation. With respect to this $\mathfrak{so}(3)$, the $N^2$-dimensional vector space of matrices decomposes under the adjoint action into irreducible representations of spin $0, 1, 2, \ldots, N-1$. Therefore, there exists a basis of $N \times N$ matrices $\left\{T^l_m\right\}, l=0,\ldots,N-1, m=-l,\ldots,l$ such that the matrix multiplication can be written as
\begin{equation}
T^{l_1}_{m_1} \star T^{l_2}_{m_2} = \sum_{|l_1-l_2| \leq l \leq l_1+l_2} C^{l_1 l_2 l}_{m_1 m_2}(N) T^l_{m_1+m_2}.
\end{equation}
Choosing the normalization of $T^l_m$ in a suitable way, we find a basis of the algebra of $N \times N$ matrices where the structure constants $C^{l_1 l_2 l}_{m_1 m_2}$ depend \emph{rationally} on $N$. Since the associativity conditions are algebraic conditions on the structure constants, we can consider the elements $T^l_m$ with $l$ running over all non-negative integers and define the $\star$ product by the same formula but without any integrality restriction on the allowed values of $N$. The associativity conditions are automatically satisfied. The rank parameter $N$ is usually denoted by $\lambda$ when it takes arbitrary (not necessarily integer) value. We therefore find a one-parametric family of associative algebras $\mathfrak{hs}(\lambda)$\footnote{Sometimes in higher spin literature $\mathfrak{hs}(\lambda)$ denotes the Lie algebra induced from the associative $\star$ multiplication. In this Lie algebra setting, the central generator $T^0_0$ decouples from $T^l_m, l \geq 1$ and $\mathfrak{hs}(\lambda)$ is often considered to be the Lie algebra with generators $T^l_m$ with $l \geq 1$.}. These algebras interpolate between algebras of $N \times N$ matrices in the following sense: restricting $\lambda$ to be a positive integer $N$, the generators $T^l_m$ with $l \geq N$ form an ideal in $\mathfrak{hs}(N)$. Quotienting out by this ideal we recover the algebra of $N \times N$ matrices.

It is interesting that in the limit $\lambda \to \infty$ the algebra $\mathfrak{hs}(\lambda)$ is the answer to the problem of deformation quantization of the Poisson algebra of functions on the two-sphere $S^2$. More concretely, since the round two-sphere has a natural rotation-invariant symplectic form (the round volume two-form), we can use it to define a Poisson bracket. The action of the Lie algebra of the rotation group $\mathfrak{so}(3)$ on the space of functions on $S^2$ determines a natural basis of spherical harmonics $Y^l_m(\vartheta,\varphi)$. The philosophy of deformation quantization instructs us to look for an associative multiplication of the form
\begin{equation}
Y^{l_1}_{m_1} \star Y^{l_2}_{m_2} = Y^{l_1}_{m_1} \cdot Y^{l_2}_{m_2} + \hbar \left\{ Y^{l_1}_{m_1}, Y^{l_2}_{m_2} \right\} + \mathcal{O}(\hbar^2).
\end{equation}
The first term is the usual pointwise multiplication of functions on $S^2$ while the next term is the Poisson bracket. By studying the large $\lambda$ expansion of $C^{l_1 l_2 l}_{m_1 m_2}(\lambda)$, we find that the $\star$ product in $\mathfrak{hs}(\lambda)$ is indeed of this form with $\hbar \sim \lambda^{-1}$. Therefore, the associative algebra $\mathfrak{hs}(\lambda)$ interpolates between the algebras of matrix multiplication for $\lambda \in \mathbbm{N}$ while at $\lambda \to \infty$ it reduces to algebra of functions on $S^2$ with pointwise multiplication and the natural Poisson bracket.

The previous discussion can be understood in a more abstract algebraic way as follows: the associative algebra $\mathfrak{hs}(\lambda)$ can be defined as a quotient of the universal enveloping algebra of $\mathfrak{so}(3)$
\begin{equation}
\mathfrak{hs}(\lambda) \simeq \frac{\mathcal{U}(\mathfrak{so}(3))}{\langle X^2+Y^2+Z^2-(\lambda^2-1)\rangle}.
\end{equation}
Given an $N$-dimensional irreducible representation of $\mathfrak{so}(3)$, we have a natural map from $\mathcal{U}(\mathfrak{so}(3))$ to the associative algebra of $N \times N$ matrices. The kernel of this map is an ideal in $\mathcal{U}(\mathfrak{so}(3))$ which contains the Casimir constraint, representing the value of the quadratic Casimir element in $N$ dimesional irreducile representation. Therefore, for $\lambda = N$ we have a well-defined map from $\mathfrak{hs}(\lambda)$ to the associative algebra of $N \times N$ matrices.

Geometrically, we can think of the universal enveloping algebra $\mathcal{U}(\mathfrak{so}(3))$ to be a non-commutative generalization of $\mathbbm{R}^3$ where the coordinates $X$, $Y$ and $Z$ no longer commute but instead satisfy relation $\left[ X, Y \right] = Z$ and its cyclic permutations. The Casimir constraint
\begin{equation}
X^2 + Y^2 + Z^2 = \lambda^2-1
\end{equation}
means that we are restricting to a non-commutative (fuzzy) sphere $S^2$ inside of this $\mathbbm{R}^3$. The algebra of functions on this fuzzy sphere is exactly $\mathfrak{hs}(\lambda)$. The parameter $\lambda$ controls the area of the fuzzy sphere. For large volume, the fuzzy sphere is approaching the classical geometric two-sphere. On the other hand, when the area is quantized such that $\lambda$ is an integer, the associated quantum Hilbert space becomes finite dimensional and the corresponding operators reduce to $N \times N$ matrices.

\subsubsection{$\mathcal{W}_\infty$}

Parallel to the simpler situation of $\mathfrak{hs}(\lambda)$, we can ask if there is an analogous algebra that would interpolate between all algebras of the $\mathcal{W}_N$ family. Such an algebra was studied in \cite{Gaberdiel:2012ku} and was called $\mathcal{W}_\infty[\lambda,c]$\footnote{In the older literature such as \cite{Pope:1989ew}, one-parametric families of algebras $\mathcal{W}_\infty$ and $\mathcal{W}_{1+\infty}$ were studied. The authors essentially took the $N \to \infty$ limit of non-linear $\mathcal{W}_N$ algebras and obtained a family of algebras that had one less parameter but had the advantage of being \emph{linear}, i.e. Lie algebras. It took more than 20 years until \cite{Gaberdiel:2012ku} realized the importance of full two-parametric family of $\mathcal{W}_\infty$ algebras, despite these being non-linear in general.}. We can either apply the Zamolodchikov-style bootstrap to find algebras with generating fields of spin $2, 3, \ldots$ with no upper bound on the dimension of the generating fields \cite{Linshaw:2017tvv}, or alternatively we can apply one of the constructions of $\mathcal{W}_N$ algebras (GKO coset or Drinfeld-Sokolov reduction) directly to $\mathfrak{hs}(\lambda)$. Both ways lead to $\mathcal{W}_\infty[\lambda,c]$ and the resulting algebra has the interpolating property analogous to $\mathfrak{hs}(\lambda)$: choosing $\lambda$ to be a positive integer $N$, the generating fields of dimension $N+1$ and higher form an ideal in $\mathcal{W}_\infty$ and quotienting out by this ideal results in $\mathcal{W}_N[c]$. Practically, in order to study $\mathcal{W}_N$ algebras, it does not make much difference whether we study simultaneously $\mathcal{W}_N$ for all integer ranks or if we study the family $\mathcal{W}_\infty[\lambda,c]$ with generic parameter $\lambda$. The main property of the family is that with suitable choice of normalization of the generators the structure constants of the algebras are algebraic functions of $N$ or $\lambda$.

Apart from $\mathcal{W}_\infty$ there is also a closely related algebra $\mathcal{W}_{1+\infty} \simeq \widehat{\mathfrak{u}}(1) \times \mathcal{W}_\infty$, i.e. a product of $\mathcal{W}_\infty$ and a decoupled free boson. These two algebras are related in the analogous way as $\mathfrak{sl}(N)$ and $\mathfrak{gl}(N)$. For some purposes it is more convenient to consider the simple algebra, i.e. $\mathcal{W}_\infty$, but as we will see in the following, for other purposes it turns out to be extremely useful to include the additional decoupled spin $1$ generator.

\subsubsection{Triality}
One of the most important observations made by Gaberdiel and Gopakumar in \cite{Gaberdiel:2012ku} about $\mathcal{W}_\infty$ was that at fixed central charge $c$ the structure constants of $\mathcal{W}_\infty$ depend on the rank parameter $\lambda$ via a degree $3$ rational function, i.e. that for a fixed value of the central charge $c$ there are three choices of the parameter $\lambda$ that have identical OPEs. Denoting these three values by $\lambda_1, \lambda_2$ and $\lambda_3$, we have
\begin{equation}
\mathcal{W}_\infty[\lambda_1,c] \simeq \mathcal{W}_\infty[\lambda_2,c] \simeq \mathcal{W}_\infty[\lambda_3,c]
\end{equation}
where the three values of $\lambda$ are constrained by
\begin{equation}
\label{lambdarestriction}
\frac{1}{\lambda_1} + \frac{1}{\lambda_2} + \frac{1}{\lambda_3} = 0
\end{equation}
and the central charge (in $\mathcal{W}_\infty$) is
\begin{equation}
c = (\lambda_1-1)(\lambda_2-1)(\lambda_3-1).
\end{equation}
The authors called the related $\mathcal{S}_3$ symmetry permuting the $\lambda$-parameters the \emph{triality} symmetry. This symmetry underlies most of the representation theory of the algebra as we will see in the following.

\subsubsection{Vacuum representation}

\begin{figure}[t]
\sidecaption
\planepartition{{4,3,2,1},{3,3,1},{3,2},{1}}
\caption{An example of plane partition with $23$ boxes representing a state in the vacuum representation.}
\label{planepartition}
\end{figure}


As an example of how the triality symmetry underlies the representation theory of $\mathcal{W}_{1+\infty}$, consider the vacuum representation of $\mathcal{W}_{1+\infty}$. The vectors in the vacuum representation are obtained by acting on the highest weight state $\ket{0}$ by $J_{-1}, J_{-2}, J_{-3}, \ldots$, $L_{-2}, L_{-3}, \ldots$, $W_{-3}, \ldots$ where $J(z),T(z),W(z),\ldots$ are generating fields of spin $1, 2, 3, \ldots$ and $J_m, L_m, W_m,\ldots$ are their Fourier modes. In total, we have one letter $J_{-1}$ of mode number $1$, two letters $J_{-2}$ and $L_{-2}$ of mode number $2$, three letters of mode number $3$ etc. The total vacuum character which we can think of as counting all the local fields in $\mathcal{W}_{1+\infty}$ algebra is thus given by
\begin{equation}
\label{macmahon}
\prod_{n=1}^\infty \frac{1}{(1-q^n)^n} \simeq 1 + q + 3q^2 + 6q^3 + 13q^4 + 24q^5 + 48q^6 + \ldots.
\end{equation}
Note the similarity with the counting of the states of the Fock space of a single free boson. In that situation we have one letter of every mode number and the resulting character is counting partitions of integers or in other words Young diagrams.

It is a rather non-trivial result that the function \eqref{macmahon}, the MacMahon function, is counting three-dimensional generalization of the Young diagrams, so-called \emph{plane partitions} (see Figure \ref{planepartition}). We come to an interesting conclusion, namely that the counting problem of the local fields in $\mathcal{W}_{1+\infty}$ algebra is the same as the problem of counting the number of possible ways of stacking the boxes in the corner of a room following the plane partition rule: no box can be continuously moved in the direction of the corner, i.e. every box has to be supported from the three directions towards the corner by other boxes or by the walls.

There is a natural candidate for the action of $\mathcal{S}_3$ triality symmetry on the space of plane partitions: the action just corresponds to the permutation of the coordinate axes. Another property of $\mathcal{W}_{1+\infty}$ that can be nicely seen from the plane partition picture is the truncation $\mathcal{W}_\infty \to \mathcal{W}_N$. If we specialize $\lambda \to N$ and truncate the algebra, the corresponding vacuum character counts configurations of boxes where there are only $N$ layers allowed in one of the three directions.

\subsubsection{Truncations of the algebra and minimal models}

\begin{figure}[t]
\sidecaption
\boxed{ \includegraphics[scale=0.50]{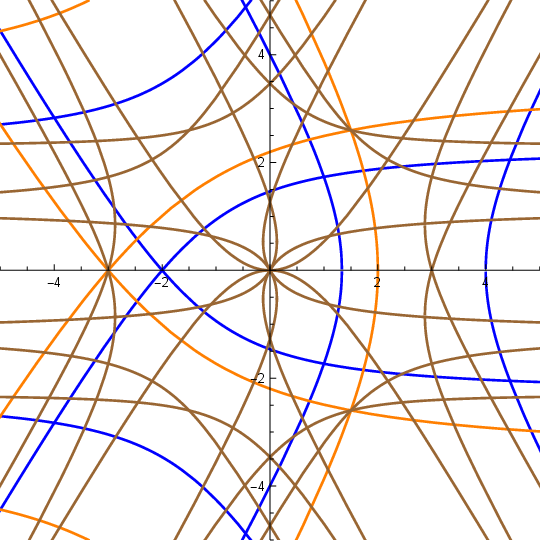}}
\caption{Truncation curves of $\mathcal{W}_{\infty}$ algebra. Every curve corresponds to locus in the parameter space where the vacuum Verma module is reducible. Double intersections correspond to minimal models of $\mathcal{W}_\infty$. The triality symmetry is manifest in this diagram.}
\label{figtrunccurves}
\end{figure}

Another place where the hidden triality symmetry nicely shows up is in the context of truncations of $\mathcal{W}_\infty$ \cite{Prochazka:2014gqa}. From the construction of $\mathcal{W}_\infty$ we already know that after a specialization of $\lambda_j$ to a positive integer, the vacuum Verma module of $\mathcal{W}_\infty$ is no longer irreducible and the corresponding irreducible quotient reduces to $\mathcal{W}_N$ algebra. We can ask if there are any other specializations of $\lambda$-parameters where the vacuum representation becomes reducible. Analogously to the situation in Virasoro algebra, one can calculate the matrix of all inner product of states in vacuum representation of the algebra. Its determinant, the Kac determinant, which is a function of the Virasoro level and the $\lambda$-parameters, vanishes precisely when there is such a reduction. In \cite{Prochazka:2014gqa} these Kac determinants were studied up to level $10$ and the following condition was found (see also \cite{Prochazka:2017qum}): if the $\lambda_j$ parameters satisfy the equation
\begin{equation}
\label{trunccurves}
\frac{N_1}{\lambda_1} + \frac{N_2}{\lambda_2} + \frac{N_3}{\lambda_3} = 1
\end{equation}
with $N_j$ non-negative integers, the vacuum Verma module of the algebra $\mathcal{W}_{1+\infty}$ has a null state at level
\begin{equation}
\label{truncnullstate}
(N_1+1)(N_2+1)(N_3+1).
\end{equation}
The corresponding quotients were identified in \cite{Prochazka:2017qum} with $Y_{N_1 N_2 N_3}$ algebras constructed independently by Gaiotto and Rap\v{c}\'{a}k \cite{Gaiotto:2017euk} from the perspective of four-dimensional twisted maximally supersymmetric Yang-Mills theory (see also \cite{Creutzig:2020zaj}).

The curves \eqref{trunccurves} in the $\lambda$-parameter space look as in Figure \ref{figtrunccurves}. Each curve is a codimension $1$ specialization of the two-parametric family $\mathcal{W}_\infty$. The $\mathcal{W}_N$ algebras correspond to specializations of the form $(N_1,N_2,N_3)=(0,0,N)$. The other more general specializations correspond to various minimal models: for instance the choice $(N_1,N_2,N_3)=(2,1,0)$ corresponds to so-called parafermions. Imposing simultaneously the parafermion constraint as well as the $\mathcal{W}_N$ condition
\begin{equation}
\frac{2}{\lambda_1} + \frac{1}{\lambda_2} = 1, \qquad \frac{N}{\lambda_3} = 1
\end{equation}
together with \eqref{lambdarestriction} determines all three $\lambda$-parameters and determines the central charge to be
\begin{equation}
c = \frac{2(N-1)}{N+2}
\end{equation}
which exactly corresponds to the series of first unitary minimal models of $\mathcal{W}_N$ algebras. For $N=2$ we get the Ising model with $c = \frac{1}{2}$ while for $\mathcal{W}_3$ we find the minimal model with $c = \frac{4}{5}$. All the first unitary minimal models have a null state at level $6$ in the vacuum representation which is exactly what the formula \eqref{truncnullstate} predicts in this simple situation \cite{Prochazka:2018tlo}.

\subsubsection{Other representations}

\begin{figure}[t]
\sidecaption
\planepartitionsm{{16,16,16,8,6,6,5,3,3,3,3,3,3,3,3,3},{16,12,9,5,5,3,2,2,2,1,1,1,1,1,1,1},{16,7,4,2,2,2,1,1},{11,6,3,2,2,2,1},{7,4,2,1,1},{6,3,2,1},{4,3,2},{4,3,2},{3,2,2},{2,2,1},{2,2,1},{2,2,1},{2,2,1},{2,2,1},{2,2,1},{2,2,1}}
\caption{A typical plane partition with non-trivial Young diagram asymptotics along the coordinate directions corresponding to a non-trivial primary.}
\label{planepartitionasym}
\end{figure}

We can also study other highest weight representations. A rather large class of representations corresponds to box counting with non-trivial Young-diagram asymptotics along the three coordinate axes (see Figure \ref{planepartitionasym}). In particular, the so-called maximally degenerate representations of $\mathcal{W}_N$ algebras are labeled by a pair of Young diagram and the characters of these representations for a generic central charge precisely correspond to counting of boxes as in Figure \ref{planepartitionasym} (but with only two non-trivial Young diagram asymptotics along two axes and bound $N$ on the height of boxes in the third direction). Interestingly, the same box counting appears in the context of the topological string theory in the context of the \emph{topological vertex}. \cite{Aganagic:2003db,Prochazka:2015deb}.

\begin{figure}
\sidecaption
\begin{tikzpicture}[scale=0.25]
\definecolor{mbr}{RGB}{128,71,0}
\definecolor{mgr}{RGB}{148,166,0}
\definecolor{mor}{RGB}{242,198,0}
\filldraw[fill=mor,draw=black] (30:1) -- ++(210:1) -- ++(330:1) -- ++(30:1) -- ++(150:1);
\filldraw[fill=mor,draw=black] (330:1) ++(30:1) -- ++(210:1) -- ++(330:1) -- ++(30:1) -- ++(150:1);
\filldraw[fill=mor,draw=black] (330:2) ++(30:1) -- ++(210:1) -- ++(330:1) -- ++(30:1) -- ++(150:1);
\filldraw[fill=mor,draw=black] (330:3) ++(30:1) -- ++(210:1) -- ++(330:1) -- ++(30:1) -- ++(150:1);
\filldraw[fill=mor,draw=black] (330:4) ++(30:1) -- ++(210:1) -- ++(330:1) -- ++(30:1) -- ++(150:1);
\filldraw[fill=mor,draw=black] (330:5) ++(30:1) -- ++(210:1) -- ++(330:1) -- ++(30:1) -- ++(150:1);
\filldraw[fill=mor,draw=black] (330:6) ++(30:1) -- ++(210:1) -- ++(330:1) -- ++(30:1) -- ++(150:1);
\filldraw[fill=mor,draw=black] (330:7) ++(30:1) -- ++(210:1) -- ++(330:1) -- ++(30:1) -- ++(150:1);
\filldraw[fill=mor,draw=black] (330:8) ++(30:1) -- ++(210:1) -- ++(330:1) -- ++(30:1) -- ++(150:1);
\filldraw[fill=mor,draw=black] (330:9) ++(30:1) -- ++(210:1) -- ++(330:1) -- ++(30:1) -- ++(150:1);
\filldraw[fill=mor,draw=black] (210:1) ++(30:1) -- ++(210:1) -- ++(330:1) -- ++(30:1) -- ++(150:1);
\filldraw[fill=mor,draw=black] (210:2) ++(30:1) -- ++(210:1) -- ++(330:1) -- ++(30:1) -- ++(150:1);
\filldraw[fill=mor,draw=black] (210:3) ++(30:1) -- ++(210:1) -- ++(330:1) -- ++(30:1) -- ++(150:1);
\filldraw[fill=mor,draw=black] (210:4) ++(30:1) -- ++(210:1) -- ++(330:1) -- ++(30:1) -- ++(150:1);
\filldraw[fill=mor,draw=black] (210:5) ++(30:1) -- ++(210:1) -- ++(330:1) -- ++(30:1) -- ++(150:1);
\filldraw[fill=mor,draw=black] (210:6) ++(30:1) -- ++(210:1) -- ++(330:1) -- ++(30:1) -- ++(150:1);
\filldraw[fill=mor,draw=black] (210:7) ++(30:1) -- ++(210:1) -- ++(330:1) -- ++(30:1) -- ++(150:1);
\filldraw[fill=mor,draw=black] (210:8) ++(30:1) -- ++(210:1) -- ++(330:1) -- ++(30:1) -- ++(150:1);
\filldraw[fill=mor,draw=black] (210:9) ++(30:1) -- ++(210:1) -- ++(330:1) -- ++(30:1) -- ++(150:1);
\filldraw[fill=mor,draw=black] (330:1) ++(210:1) ++(30:1) -- ++(210:1) -- ++(330:1) -- ++(30:1) -- ++(150:1);
\filldraw[fill=mor,draw=black] (330:1) ++(210:2) ++(30:1) -- ++(210:1) -- ++(330:1) -- ++(30:1) -- ++(150:1);
\filldraw[fill=mor,draw=black] (330:1) ++(210:3) ++(30:1) -- ++(210:1) -- ++(330:1) -- ++(30:1) -- ++(150:1);
\filldraw[fill=mor,draw=black] (330:1) ++(210:4) ++(30:1) -- ++(210:1) -- ++(330:1) -- ++(30:1) -- ++(150:1);
\filldraw[fill=mor,draw=black] (330:1) ++(210:5) ++(30:1) -- ++(210:1) -- ++(330:1) -- ++(30:1) -- ++(150:1);
\filldraw[fill=mor,draw=black] (330:1) ++(210:6) ++(30:1) -- ++(210:1) -- ++(330:1) -- ++(30:1) -- ++(150:1);
\filldraw[fill=mor,draw=black] (330:1) ++(210:7) ++(30:1) -- ++(210:1) -- ++(330:1) -- ++(30:1) -- ++(150:1);
\filldraw[fill=mor,draw=black] (330:1) ++(210:8) ++(30:1) -- ++(210:1) -- ++(330:1) -- ++(30:1) -- ++(150:1);
\filldraw[fill=mor,draw=black] (330:1) ++(210:9) ++(30:1) -- ++(210:1) -- ++(330:1) -- ++(30:1) -- ++(150:1);
\filldraw[fill=mgr,draw=black] (210:1) ++(330:3) -- ++(90:1) -- ++(210:1) -- ++(270:1) -- ++(30:1);
\filldraw[fill=mgr,draw=black] (210:2) ++(330:3) -- ++(90:1) -- ++(210:1) -- ++(270:1) -- ++(30:1);
\filldraw[fill=mgr,draw=black] (210:3) ++(330:3) -- ++(90:1) -- ++(210:1) -- ++(270:1) -- ++(30:1);
\filldraw[fill=mgr,draw=black] (210:4) ++(330:3) -- ++(90:1) -- ++(210:1) -- ++(270:1) -- ++(30:1);
\filldraw[fill=mgr,draw=black] (210:5) ++(330:3) -- ++(90:1) -- ++(210:1) -- ++(270:1) -- ++(30:1);
\filldraw[fill=mgr,draw=black] (210:6) ++(330:3) -- ++(90:1) -- ++(210:1) -- ++(270:1) -- ++(30:1);
\filldraw[fill=mgr,draw=black] (210:7) ++(330:3) -- ++(90:1) -- ++(210:1) -- ++(270:1) -- ++(30:1);
\filldraw[fill=mgr,draw=black] (210:8) ++(330:3) -- ++(90:1) -- ++(210:1) -- ++(270:1) -- ++(30:1);
\filldraw[fill=mgr,draw=black] (210:9) ++(330:3) -- ++(90:1) -- ++(210:1) -- ++(270:1) -- ++(30:1);
\filldraw[fill=mgr,draw=black] (210:2) ++(330:4) -- ++(90:1) -- ++(210:1) -- ++(270:1) -- ++(30:1);
\filldraw[fill=mgr,draw=black] (210:3) ++(330:4) -- ++(90:1) -- ++(210:1) -- ++(270:1) -- ++(30:1);
\filldraw[fill=mgr,draw=black] (210:4) ++(330:4) -- ++(90:1) -- ++(210:1) -- ++(270:1) -- ++(30:1);
\filldraw[fill=mgr,draw=black] (210:5) ++(330:4) -- ++(90:1) -- ++(210:1) -- ++(270:1) -- ++(30:1);
\filldraw[fill=mgr,draw=black] (210:6) ++(330:4) -- ++(90:1) -- ++(210:1) -- ++(270:1) -- ++(30:1);
\filldraw[fill=mgr,draw=black] (210:7) ++(330:4) -- ++(90:1) -- ++(210:1) -- ++(270:1) -- ++(30:1);
\filldraw[fill=mgr,draw=black] (210:8) ++(330:4) -- ++(90:1) -- ++(210:1) -- ++(270:1) -- ++(30:1);
\filldraw[fill=mgr,draw=black] (210:9) ++(330:4) -- ++(90:1) -- ++(210:1) -- ++(270:1) -- ++(30:1);
\filldraw[fill=mgr,draw=black] (210:10) ++(330:4) -- ++(90:1) -- ++(210:1) -- ++(270:1) -- ++(30:1);
\filldraw[fill=mbr,draw=black] (330:2) ++(270:1) -- ++(330:1) -- ++(90:1) -- ++(150:1) -- ++(270:1);
\filldraw[fill=mbr,draw=black] (330:3) ++(270:1) -- ++(330:1) -- ++(90:1) -- ++(150:1) -- ++(270:1);
\filldraw[fill=mbr,draw=black] (330:4) ++(270:1) -- ++(330:1) -- ++(90:1) -- ++(150:1) -- ++(270:1);
\filldraw[fill=mbr,draw=black] (330:5) ++(270:1) -- ++(330:1) -- ++(90:1) -- ++(150:1) -- ++(270:1);
\filldraw[fill=mbr,draw=black] (330:6) ++(270:1) -- ++(330:1) -- ++(90:1) -- ++(150:1) -- ++(270:1);
\filldraw[fill=mbr,draw=black] (330:7) ++(270:1) -- ++(330:1) -- ++(90:1) -- ++(150:1) -- ++(270:1);
\filldraw[fill=mbr,draw=black] (330:8) ++(270:1) -- ++(330:1) -- ++(90:1) -- ++(150:1) -- ++(270:1);
\filldraw[fill=mbr,draw=black] (330:9) ++(270:1) -- ++(330:1) -- ++(90:1) -- ++(150:1) -- ++(270:1);
\filldraw[fill=mbr,draw=black] (330:2) ++(270:2) -- ++(330:1) -- ++(90:1) -- ++(150:1) -- ++(270:1);
\filldraw[fill=mbr,draw=black] (330:3) ++(270:2) -- ++(330:1) -- ++(90:1) -- ++(150:1) -- ++(270:1);
\filldraw[fill=mbr,draw=black] (330:4) ++(270:2) -- ++(330:1) -- ++(90:1) -- ++(150:1) -- ++(270:1);
\filldraw[fill=mbr,draw=black] (330:5) ++(270:2) -- ++(330:1) -- ++(90:1) -- ++(150:1) -- ++(270:1);
\filldraw[fill=mbr,draw=black] (330:6) ++(270:2) -- ++(330:1) -- ++(90:1) -- ++(150:1) -- ++(270:1);
\filldraw[fill=mbr,draw=black] (330:7) ++(270:2) -- ++(330:1) -- ++(90:1) -- ++(150:1) -- ++(270:1);
\filldraw[fill=mbr,draw=black] (330:8) ++(270:2) -- ++(330:1) -- ++(90:1) -- ++(150:1) -- ++(270:1);
\filldraw[fill=mbr,draw=black] (330:9) ++(270:2) -- ++(330:1) -- ++(90:1) -- ++(150:1) -- ++(270:1);
\filldraw[fill=mor,draw=black] (270:4) ++ (30:2) -- ++(210:1) -- ++(330:1) -- ++(30:1) -- ++(150:1);
\filldraw[fill=mor,draw=black] (270:4) ++ (30:2) ++ (330:1) -- ++(210:1) -- ++(330:1) -- ++(30:1) -- ++(150:1);
\filldraw[fill=mor,draw=black] (270:4) ++ (30:2) ++ (330:2) -- ++(210:1) -- ++(330:1) -- ++(30:1) -- ++(150:1);
\filldraw[fill=mor,draw=black] (270:4) ++ (30:2) ++ (330:3) -- ++(210:1) -- ++(330:1) -- ++(30:1) -- ++(150:1);
\filldraw[fill=mor,draw=black] (270:4) ++ (30:2) ++ (330:4) -- ++(210:1) -- ++(330:1) -- ++(30:1) -- ++(150:1);
\filldraw[fill=mor,draw=black] (270:4) ++ (30:2) ++ (330:5) -- ++(210:1) -- ++(330:1) -- ++(30:1) -- ++(150:1);
\filldraw[fill=mor,draw=black] (270:4) ++ (30:2) ++ (330:6) -- ++(210:1) -- ++(330:1) -- ++(30:1) -- ++(150:1);
\filldraw[fill=mor,draw=black] (270:4) ++ (30:2) ++ (330:7) -- ++(210:1) -- ++(330:1) -- ++(30:1) -- ++(150:1);
\filldraw[fill=mor,draw=black] (270:4) ++ (30:2) ++ (210:1) -- ++(210:1) -- ++(330:1) -- ++(30:1) -- ++(150:1);
\filldraw[fill=mor,draw=black] (270:4) ++ (30:2) ++ (210:2) -- ++(210:1) -- ++(330:1) -- ++(30:1) -- ++(150:1);
\filldraw[fill=mor,draw=black] (270:4) ++ (30:2) ++ (210:3) -- ++(210:1) -- ++(330:1) -- ++(30:1) -- ++(150:1);
\filldraw[fill=mor,draw=black] (270:4) ++ (30:2) ++ (210:4) -- ++(210:1) -- ++(330:1) -- ++(30:1) -- ++(150:1);
\filldraw[fill=mor,draw=black] (270:4) ++ (30:2) ++ (210:5) -- ++(210:1) -- ++(330:1) -- ++(30:1) -- ++(150:1);
\filldraw[fill=mor,draw=black] (270:4) ++ (30:2) ++ (210:6) -- ++(210:1) -- ++(330:1) -- ++(30:1) -- ++(150:1);
\filldraw[fill=mor,draw=black] (270:4) ++ (30:2) ++ (210:7) -- ++(210:1) -- ++(330:1) -- ++(30:1) -- ++(150:1);
\filldraw[fill=mor,draw=black] (270:4) ++ (30:2) ++ (210:8) -- ++(210:1) -- ++(330:1) -- ++(30:1) -- ++(150:1);
\filldraw[fill=mor,draw=black] (270:4) ++ (330:1) ++(210:-1) -- ++(210:1) -- ++(330:1) -- ++(30:1) -- ++(150:1);
\filldraw[fill=mor,draw=black] (270:4) ++ (330:1) ++(210:0) -- ++(210:1) -- ++(330:1) -- ++(30:1) -- ++(150:1);
\filldraw[fill=mor,draw=black] (270:4) ++ (330:1) ++(210:1) -- ++(210:1) -- ++(330:1) -- ++(30:1) -- ++(150:1);
\filldraw[fill=mor,draw=black] (270:4) ++ (330:1) ++(210:2) -- ++(210:1) -- ++(330:1) -- ++(30:1) -- ++(150:1);
\filldraw[fill=mor,draw=black] (270:4) ++ (330:1) ++(210:3) -- ++(210:1) -- ++(330:1) -- ++(30:1) -- ++(150:1);
\filldraw[fill=mor,draw=black] (270:4) ++ (330:1) ++(210:4) -- ++(210:1) -- ++(330:1) -- ++(30:1) -- ++(150:1);
\filldraw[fill=mor,draw=black] (270:4) ++ (330:1) ++(210:5) -- ++(210:1) -- ++(330:1) -- ++(30:1) -- ++(150:1);
\filldraw[fill=mor,draw=black] (270:4) ++ (330:1) ++(210:6) -- ++(210:1) -- ++(330:1) -- ++(30:1) -- ++(150:1);
\filldraw[fill=mgr,draw=black] (210:4) ++(330:7) -- ++(90:1) -- ++(210:1) -- ++(270:1) -- ++(30:1);
\filldraw[fill=mgr,draw=black] (210:5) ++(330:7) -- ++(90:1) -- ++(210:1) -- ++(270:1) -- ++(30:1);
\filldraw[fill=mgr,draw=black] (210:6) ++(330:7) -- ++(90:1) -- ++(210:1) -- ++(270:1) -- ++(30:1);
\filldraw[fill=mgr,draw=black] (210:7) ++(330:7) -- ++(90:1) -- ++(210:1) -- ++(270:1) -- ++(30:1);
\filldraw[fill=mgr,draw=black] (210:8) ++(330:7) -- ++(90:1) -- ++(210:1) -- ++(270:1) -- ++(30:1);
\filldraw[fill=mgr,draw=black] (210:9) ++(330:7) -- ++(90:1) -- ++(210:1) -- ++(270:1) -- ++(30:1);
\filldraw[fill=mgr,draw=black] (210:10) ++(330:7) -- ++(90:1) -- ++(210:1) -- ++(270:1) -- ++(30:1);
\filldraw[fill=mgr,draw=black] (210:11) ++(330:7) -- ++(90:1) -- ++(210:1) -- ++(270:1) -- ++(30:1);
\filldraw[fill=mgr,draw=black] (210:5) ++(330:8) -- ++(90:1) -- ++(210:1) -- ++(270:1) -- ++(30:1);
\filldraw[fill=mgr,draw=black] (210:6) ++(330:8) -- ++(90:1) -- ++(210:1) -- ++(270:1) -- ++(30:1);
\filldraw[fill=mgr,draw=black] (210:7) ++(330:8) -- ++(90:1) -- ++(210:1) -- ++(270:1) -- ++(30:1);
\filldraw[fill=mgr,draw=black] (210:8) ++(330:8) -- ++(90:1) -- ++(210:1) -- ++(270:1) -- ++(30:1);
\filldraw[fill=mgr,draw=black] (210:9) ++(330:8) -- ++(90:1) -- ++(210:1) -- ++(270:1) -- ++(30:1);
\filldraw[fill=mgr,draw=black] (210:10) ++(330:8) -- ++(90:1) -- ++(210:1) -- ++(270:1) -- ++(30:1);
\filldraw[fill=mgr,draw=black] (210:11) ++(330:8) -- ++(90:1) -- ++(210:1) -- ++(270:1) -- ++(30:1);
\filldraw[fill=mgr,draw=black] (210:12) ++(330:8) -- ++(90:1) -- ++(210:1) -- ++(270:1) -- ++(30:1);
\filldraw[fill=mbr,draw=black] (330:3) ++(270:4) -- ++(330:1) -- ++(90:1) -- ++(150:1) -- ++(270:1);
\filldraw[fill=mbr,draw=black] (330:4) ++(270:4) -- ++(330:1) -- ++(90:1) -- ++(150:1) -- ++(270:1);
\filldraw[fill=mbr,draw=black] (330:5) ++(270:4) -- ++(330:1) -- ++(90:1) -- ++(150:1) -- ++(270:1);
\filldraw[fill=mbr,draw=black] (330:6) ++(270:4) -- ++(330:1) -- ++(90:1) -- ++(150:1) -- ++(270:1);
\filldraw[fill=mbr,draw=black] (330:7) ++(270:4) -- ++(330:1) -- ++(90:1) -- ++(150:1) -- ++(270:1);
\filldraw[fill=mbr,draw=black] (330:8) ++(270:4) -- ++(330:1) -- ++(90:1) -- ++(150:1) -- ++(270:1);
\filldraw[fill=mbr,draw=black] (330:3) ++(270:5) -- ++(330:1) -- ++(90:1) -- ++(150:1) -- ++(270:1);
\filldraw[fill=mbr,draw=black] (330:4) ++(270:5) -- ++(330:1) -- ++(90:1) -- ++(150:1) -- ++(270:1);
\filldraw[fill=mbr,draw=black] (330:5) ++(270:5) -- ++(330:1) -- ++(90:1) -- ++(150:1) -- ++(270:1);
\filldraw[fill=mbr,draw=black] (330:6) ++(270:5) -- ++(330:1) -- ++(90:1) -- ++(150:1) -- ++(270:1);
\filldraw[fill=mbr,draw=black] (330:7) ++(270:5) -- ++(330:1) -- ++(90:1) -- ++(150:1) -- ++(270:1);
\filldraw[fill=mbr,draw=black] (330:8) ++(270:5) -- ++(330:1) -- ++(90:1) -- ++(150:1) -- ++(270:1);
\filldraw[fill=mor,draw=black] (270:8) ++ (30:3) -- ++(210:1) -- ++(330:1) -- ++(30:1) -- ++(150:1);
\filldraw[fill=mor,draw=black] (270:8) ++ (30:3) ++ (330:1) -- ++(210:1) -- ++(330:1) -- ++(30:1) -- ++(150:1);
\filldraw[fill=mor,draw=black] (270:8) ++ (30:3) ++ (330:2) -- ++(210:1) -- ++(330:1) -- ++(30:1) -- ++(150:1);
\filldraw[fill=mor,draw=black] (270:8) ++ (30:3) ++ (330:3) -- ++(210:1) -- ++(330:1) -- ++(30:1) -- ++(150:1);
\filldraw[fill=mor,draw=black] (270:8) ++ (30:3) ++ (330:4) -- ++(210:1) -- ++(330:1) -- ++(30:1) -- ++(150:1);
\filldraw[fill=mor,draw=black] (270:8) ++ (30:3) ++ (330:5) -- ++(210:1) -- ++(330:1) -- ++(30:1) -- ++(150:1);
\filldraw[fill=mor,draw=black] (270:8) ++ (30:3) ++ (210:1) -- ++(210:1) -- ++(330:1) -- ++(30:1) -- ++(150:1);
\filldraw[fill=mor,draw=black] (270:8) ++ (30:3) ++ (210:2) -- ++(210:1) -- ++(330:1) -- ++(30:1) -- ++(150:1);
\filldraw[fill=mor,draw=black] (270:8) ++ (30:3) ++ (210:3) -- ++(210:1) -- ++(330:1) -- ++(30:1) -- ++(150:1);
\filldraw[fill=mor,draw=black] (270:8) ++ (30:3) ++ (210:4) -- ++(210:1) -- ++(330:1) -- ++(30:1) -- ++(150:1);
\filldraw[fill=mor,draw=black] (270:8) ++ (30:3) ++ (210:5) -- ++(210:1) -- ++(330:1) -- ++(30:1) -- ++(150:1);
\filldraw[fill=mor,draw=black] (270:8) ++ (30:3) ++ (210:6) -- ++(210:1) -- ++(330:1) -- ++(30:1) -- ++(150:1);
\filldraw[fill=mor,draw=black] (270:8) ++ (30:3) ++ (210:7) -- ++(210:1) -- ++(330:1) -- ++(30:1) -- ++(150:1);
\filldraw[fill=mor,draw=black] (270:8) ++ (330:1) ++(210:-2) -- ++(210:1) -- ++(330:1) -- ++(30:1) -- ++(150:1);
\filldraw[fill=mor,draw=black] (270:8) ++ (330:1) ++(210:-1) -- ++(210:1) -- ++(330:1) -- ++(30:1) -- ++(150:1);
\filldraw[fill=mor,draw=black] (270:8) ++ (330:1) ++(210:0) -- ++(210:1) -- ++(330:1) -- ++(30:1) -- ++(150:1);
\filldraw[fill=mor,draw=black] (270:8) ++ (330:1) ++(210:1) -- ++(210:1) -- ++(330:1) -- ++(30:1) -- ++(150:1);
\filldraw[fill=mor,draw=black] (270:8) ++ (330:1) ++(210:2) -- ++(210:1) -- ++(330:1) -- ++(30:1) -- ++(150:1);
\filldraw[fill=mor,draw=black] (270:8) ++ (330:1) ++(210:3) -- ++(210:1) -- ++(330:1) -- ++(30:1) -- ++(150:1);
\filldraw[fill=mor,draw=black] (270:8) ++ (330:1) ++(210:4) -- ++(210:1) -- ++(330:1) -- ++(30:1) -- ++(150:1);
\end{tikzpicture}
\caption{Periodic plane partition corresponding to the vacuum (primary state with $\Delta=0$) of the Ising model, $c=\frac{1}{2}$.}
\label{isingdelta0}
\end{figure}
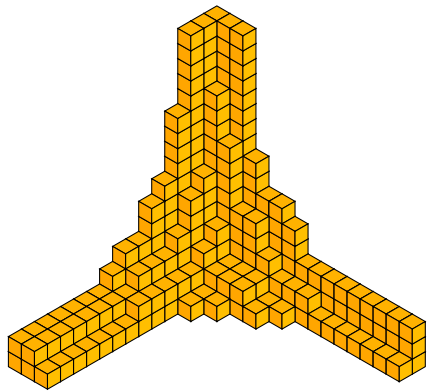

The fact that the characters of this nice class of representations of $\mathcal{W}_N$ have a combinatorial interpretation generalizes to all $\mathcal{W}_N$ minimal models (i.e. special values of the central charge). In this case, the underlying space where we build the plane partitions is perodic (compactified on a cylinder) and the box counting such as in Figure \ref{isingdelta0} correctly reproduces the characters of all $\mathcal{W}_N$ minimal model representations \cite{Foda:2015bsa,Prochazka:2023zdb}. This gives a combinatorial interpretation to formulas for these characters which are usually given in terms of ratios of theta functions \cite{Bouwknegt:1992wg}.

\section{Affine Yangian}
\label{secyangian}
There is another very different presentation of $\mathcal{W}_{1+\infty}$ which goes under the name of Yangian of $\widehat{\mathfrak{u}}(1)$ \cite{Arbesfeld:2013as,Prochazka:2015deb,Tsymbaliuk:2014fvq} and is more natural from the point of view of quantum integrability. It is also closely related to the quantum toroidal or Ding-Iohara-Miki algebra \cite{Burban:2012sch,Ding:1996mq,Feigin:2010qea,Matsuo:2023lky,Miki:2007mer}.

It can be explicitly defined as an associative algebra with generators and relations generalizing the creation and annihilation operators of the quantum mechanical harmonic oscillator. There are three sets of generators, $\psi_j$, $e_j$ and $f_j$ where the spin index $j=0,1,2,\ldots$ and the rather long list of relations is \cite{Arbesfeld:2013as,Tsymbaliuk:2014fvq}
\begin{eqnarray}
\label{exchangerelmodes}
\nonumber
0 & = & \left[ e_{j+3}, e_k \right] - 3 \left[ e_{j+2}, e_{k+1} \right] + 3\left[ e_{j+1}, e_{k+2} \right] - \left[ e_j, e_{k+3} \right] \\
\nonumber
& & + \sigma_2 \left[ e_{j+1}, e_k \right] - \sigma_2 \left[ e_j, e_{k+1} \right] - \sigma_3 \left\{ e_j, e_k \right\} \\
\nonumber
0 & = & \left[ f_{j+3}, f_k \right] - 3 \left[ f_{j+2}, f_{k+1} \right] + 3\left[ f_{j+1}, f_{k+2} \right] - \left[ f_j, f_{k+3} \right] \\
\nonumber
& & + \sigma_2 \left[ f_{j+1}, f_k \right] - \sigma_2 \left[ f_j, f_{k+1} \right] + \sigma_3 \left\{ f_j, f_k \right\} \\
0 & = & \left[ \psi_{j+3}, e_k \right] - 3 \left[ \psi_{j+2}, e_{k+1} \right] + 3\left[ \psi_{j+1}, e_{k+2} \right] - \left[ \psi_j, e_{k+3} \right] \\
\nonumber
& & + \sigma_2 \left[ \psi_{j+1}, e_k \right] - \sigma_2 \left[ \psi_j, e_{k+1} \right] - \sigma_3 \left\{ \psi_j, e_k \right\} \\
\nonumber
0 & = & \left[ \psi_{j+3}, f_k \right] - 3 \left[ \psi_{j+2}, f_{k+1} \right] + 3\left[ \psi_{j+1}, f_{k+2} \right] - \left[ \psi_j, f_{k+3} \right] \\
\nonumber
& & + \sigma_2 \left[ \psi_{j+1}, f_k \right] - \sigma_2 \left[ \psi_j, f_{k+1} \right] + \sigma_3 \left\{ \psi_j, f_k \right\} \\
\nonumber
0 & = & \left[ \psi_j, \psi_k \right] \\
\nonumber
\psi_{j+k} & = & \left[ e_j, f_k \right]
\end{eqnarray}
together with
\begin{align}
\nonumber
\left[ \psi_0, e_j \right] & = 0, & \left[ \psi_1, e_j \right] & = 0, & \left[ \psi_2, e_j \right] & = 2 e_j, \\
\left[ \psi_0, f_j \right] & = 0, & \left[ \psi_1, f_j \right] & = 0, & \left[ \psi_2, f_j \right] & = -2f_j
\end{align}
and finally the cubic Serre-like relations
\begin{align}
0 & = \mathrm{Sym}_{(j_1,j_2,j_3)} \left[ e_{j_1}, \left[ e_{j_2}, e_{j_3+1} \right] \right], & 0 & = \mathrm{Sym}_{(j_1,j_2,j_3)} \left[ f_{j_1}, \left[ f_{j_2}, f_{j_3+1} \right] \right].
\end{align}
As in the case of $\mathcal{W}_{1+\infty}$, we actually have two-parametric family of algebras parametrized by Nekrasov-like parameters $\epsilon_1, \epsilon_2$ and $\epsilon_3$ subject to relation
\begin{equation}
\epsilon_1 + \epsilon_2 + \epsilon_3 = 0.
\end{equation}
These parameters enter the defining relations of the algebra in a symmetric way via combinations
\begin{equation}
\sigma_2 \equiv \epsilon_1 \epsilon_2 + \epsilon_1 \epsilon_3 + \epsilon_2 \epsilon_3, \qquad \sigma_3 \equiv \epsilon_1 \epsilon_2 \epsilon_3.
\end{equation}
Notice that although both commutators and anti-commutators appear in the defining relations, the algebra is not a (super-)Lie algebra.

The first set of defining relations can be simplified if we introduce the generating functions (Drinfeld currents) by
\begin{align}
e(u) & = \sum_{j=0}^{\infty} \frac{e_j}{u^{j+1}}, & f(u) & = \sum_{j=0}^{\infty} \frac{f_j}{u^{j+1}}, & \psi(u) & = 1 + \sigma_3 \sum_{j=0}^{\infty} \frac{\psi_j}{u^{j+1}}.
\end{align}
in terms of which the first four exchange relations \eqref{exchangerelmodes} simplify to \footnote{We use $\sim$ instead of equal sign because there is a slight correction for generators of low spin \cite{Prochazka:2015deb,Tsymbaliuk:2014fvq}.}
\begin{align}
e(u) e(v) & \sim \varphi(u-v) e(v) e(u), & f(u) f(v) & \sim \varphi(v-u) f(v) f(u), \\
\psi(u) e(v) & \sim \varphi(u-v) e(v) \psi(u), & \psi(u) f(v) & \sim \varphi(v-u) f(v) \psi(u).
\end{align}
We introduced a rational function
\begin{equation}
\label{structurefunction}
\varphi(u) = \frac{(u+\epsilon_1)(u+\epsilon_2)(u+\epsilon_3)}{(u-\epsilon_1)(u-\epsilon_2)(u-\epsilon_3)}
\end{equation}
which is the unique \emph{structure function} of the algebra in the Yangian description. This structure function controls most of the representation theory of the algebra. The positions of zeros and poles are combinatorially related to vertices of the three-dimensional cube -- the basic box out of which the plane partitions are built. Later we will see its relation to the scattering phase in Bethe ansatz equations \cite{Nekrasov:2009rc,Prochazka:2019dvu}.

\subsection{Yangian representation theory}
One of the main advantages of the Yangian perspective is that the representation theory is very directly related to the presentation of the algebra \cite{Prochazka:2015deb,Tsymbaliuk:2014fvq}. The generators $\psi_j$, $j=0,1,2,\ldots$ all commute so we have a natural Cartan-like subalgebra. Their simultaneous eigenstates are labeled by combinatorial objects such as partitions, plane partitions (Figure \ref{planepartition}), plane partitions with Young diagram asymptotics (Figure \ref{planepartitionasym}) or by cylindrical partitions (Figure \ref{isingdelta0}).

The action of $\psi(u)$ on such an eigenstate $\ket{\Lambda}$ gives the corresponding generating function of eigenvalues,
\begin{equation}
\psi(u) \ket{\Lambda} = \psi_0(u) \prod_{\Box \in \Lambda} \varphi(u-\epsilon_{\Box}) \ket{\Lambda}.
\end{equation}
Typically the eigenvalue of $\psi(u)$ is a rational function of $u$ determined by the combinatorial data of the plane partition $\Lambda$: apart from the contribution of the highest weight state $\psi_0(u)$ (which is typically also a rational function of $u$), we have one factor $\varphi(u-\epsilon_{\Box})$ for every box in the plane partition $\Lambda$. The function $\varphi(u)$ is the structure function of the algebra \eqref{structurefunction} and its argument is shifted by $\epsilon_{\Box}$ which is the weighted geometric position of the given box,
\begin{equation}
\epsilon_{\Box} \equiv \epsilon_1 x_1(\Box) + \epsilon_2 x_2(\Box) + \epsilon_3 x_3(\Box).
\end{equation}
Later we will see that there is a correspondence between the boxes in $\Lambda$ and the Bethe roots and $\epsilon_{\Box}$ gives the numerical value of the Bethe root (in the low-temperature limit of the auxiliary space).

The action of ladder operators $e(u)$ and $f(u)$ is likewise determined in terms of the combinatorics of plane partitions, concretely
\begin{align}
e(u) \ket{\Lambda} & = \sum_{\Box \in \Lambda^+} \frac{E(\Lambda\to\Lambda+\Box)}{u-\epsilon_{\Box}} \ket{\Lambda+\Box} \\
f(u) \ket{\Lambda} & = \sum_{\Box \in \Lambda^-} \frac{F(\Lambda\to\Lambda-\Box)}{u-\epsilon_{\Box}} \ket{\Lambda-\Box},
\end{align}
i.e. the action of $e(u)$ on $\ket{\Lambda}$ gives a linear combination of states which have one additional box compatible with the plane partition rules and analogously for the box removal operator $f(u)$. The amplitudes $E(\Lambda\to\Lambda+\Box)$ and $F(\Lambda\to\Lambda-\Box)$ have certain ambiguity related to normalization of the vectors but the invariant combinations such as $E(\Lambda\to\Lambda+\Box) F(\Lambda+\Box\to\Lambda)$ are again determined in terms of the combinatorial data \cite{Prochazka:2015deb} and the structure function \eqref{structurefunction}.

\subsection{Relation between Yangian and $\mathcal{W}$-algebra description}
\begin{figure}[t]
\sidecaption
\begin{tikzpicture}[scale=0.9]
\draw[gray!30] (-2.25,0) grid[xstep=0.5, ystep=0.5]  (2.25,2.25);
\draw[draw=black,|-latex] (4,0) -- (4,2.75) node[above] {spin (Yangian)};
\draw[draw=black,latex-latex] (-2.5,-1) +(-0.5cm,0) -- (2.5,-1) -- +(0.5cm,0) node[below] {mode (CFT)};
\draw[draw=black,-latex] (0,0) -- (0,2.75) node[above] {\small $\psi$};
\draw[draw=black,-latex] (0.5,0) -- (0.5,2.75) node[above] {\small $e$};
\draw[draw=black,-latex] (-0.5,0) -- (-0.5,2.75) node[above] {\small $f$};
\draw[draw=black,latex-latex] (-2.5,0) +(-0.5cm,0) -- (2.5,0) -- +(0.5cm,0) node[right] {\small $J$};
\draw[draw=black,latex-latex] (-2.5,0.5) +(-0.5cm,0) -- (2.5,0.5) -- +(0.5cm,0) node[right] {\small $T$};
\draw[draw=black,latex-latex] (-2.5,1) +(-0.5cm,0) -- (2.5,1) -- +(0.5cm,0) node[right] {\small $W$};
\foreach \x in {-2,-1.5,...,2}
    \foreach \y in {0,0.5,...,2}
    {
    \fill (\x,\y) circle (1 pt);
    }
\end{tikzpicture}
\caption{The difference between the Yangian description and the description in terms of local fields.}
\label{yangianwalgebracomp}
\end{figure}
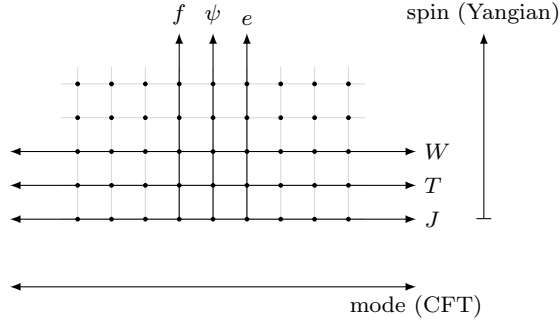

At first sight, the defining relations of the Yangian do not look anything like the operator product expansions that we use to define the $\mathcal{W}$-algebra. In fact, schematically the situation is as illustrated in Figure \ref{yangianwalgebracomp}. The Fourier modes of the $\mathcal{W}_{1+\infty}$ generators are labeled by the spin of the local field $s=1,2,\ldots$ (vertical direction of the figure) and the Fourier mode $m \in \mathbbm{Z}$ (the horizontal direction).

In the Yangian description, we focus on three Fourier modes, $m = -1, 0$ and $1$ and introduce generating fields such as $\psi(u)$ where the parameter $u$ of the generating functions (the spectral parameter) is associated to the vertical (spin) direction. On the other hand, in the usual description of $\mathcal{W}$-algebra in terms of operator product expansions we consider local fields which are the generating functions summed over all Fourier modes (horizontal direction) and the explicit OPEs typically access only lower spins. This is illustrated in the Figure \ref{yangianwalgebracomp} by spin $1$ field $J(z)$, spin $2$ field $T(z)$ and spin $3$ field $W(z)$.

As soon as we have $\sigma_3 \neq 0$, the picture becomes slightly schematic since the map between Yangian generators and Fourier modes of local fields is non-linear and non-local and also involves infinite sums of generators. For instance, for the first non-trivial generator $\psi_3$ we have the identification \cite{Prochazka:2015deb}
\begin{equation}
\label{psi3yangian}
\psi_3 \sim (W_3 + \ldots)_0 + \sigma_3 \sum_{m>0} m J_{-m} J_m.
\end{equation}
The first term on the right-hand side is a zero Fourier mode of certain local field in $\mathcal{W}_{1+\infty}$. On the other hand, the second term proportional to $\sigma_3$ involves an infinite sum of Fourier modes of spin $1$ field that \emph{can not} be written in a local form. In fact, this non-local form is characteristic of the cut-and-join operator appearing in Benjamin-Ono hierarchy of integrable equations and can only be written using a non-local operation such as the Hilbert transform \cite{Prochazka:2015deb}.

Apart from the issue just described (and the necessity to consider certain completion on the $\mathcal{W}$-algebra side), there is no problem completing the map between the two descriptions. The $\lambda$-parameters are identified as
\begin{equation}
\lambda_j = -\frac{\psi_0 \sigma_3}{\epsilon_j}
\end{equation}
and together with the identification $e_0 = J_{-1}$ and $f_0 = -J_1$ we can in principle express any Yangian generator in terms of Fourier modes of local currents and vice versa. In particular, all the infinite sums that enter the map are well-defined if we consider their actions in the highest weight representations.

\section{Integrable structures}
So far we encountered one example of what can be called the integrable structure of $\mathcal{W}_{1+\infty}$, namely the infinite set of commuting Yangian (Benjamin-Ono) generators $\psi_j$ introduced in Section \ref{secyangian}. It is rather easy to describe combinatorially their joint eigenspectrum in terms of the combinatorics of plane partitions.

On the other hand, famously there exists another choice of the commuting quantities which is natural from the $\mathcal{W}$-algebra point of view. These are the zero Fourier modes of local currents and were studied (in the Virasoro case) by Bazhanov, Lukyanov and Zamolodchikov in series of papers starting with \cite{Bazhanov:1994ft}. They are often called \emph{quantum KdV} charges and are rather difficult to explicitly diagonalize. One of the goals of this note is to explain how to connect these two choices of commuting quantities by constructing an interpolating family of \emph{quantum ILW} Hamiltonians following \cite{Litvinov:2013zda,Prochazka:2023zdb}. Before turning to that, let us shortly review the case of the classical limit, the classical KdV equation, to have at hand a classical picture of this integrable structure.

\subsection{Classical KdV}
Let us consider the classical analogue of the Virasoro stress-energy tensor $T(x)$, namely the one-dimensional Schr\"odinger operator
\begin{equation}
\label{opschroedinger}
L \equiv \partial_x^2 + u(x).
\end{equation}
Here $u(x)$ is (minus) the classical potential in which the one-dimensional particle is moving and $\partial_x^2$ represents (minus) the kinetic energy of the particle. We should specify the Hilbert space of wave functions on which the operator \eqref{opschroedinger} is acting, but as most of the discussion is purely algebraic, for our purposes the details of the corresponding function space and boundary conditions do not matter.

We can ask to what extent does the spectrum\footnote{In the case that the spectrum is not discrete one should specify what kind of spectral data we mean by \emph{spectrum} but again, details of this do not play an important role at our level of discussion.} of operator \eqref{opschroedinger} determine the potential $u(x)$. There are clearly deformations of $u(x)$ which do not change the spectrum, i.e. they are \emph{isospectral} -- the simplest example being the rigid spatial translation. This can be phrased as follows: let us continuously deform the potential according to the differential equation
\begin{equation}
\label{eqtranslation}
\partial_{t_1} u = \partial_x u.
\end{equation}
This equation is easily solved by writing $u(x,t_1) = u(x+t_1,0)$ and trivially the spectrum is independent of the deformation parameter or \emph{time} $t_1$. But perhaps surprisingly, there exist infinitely many other continuous deformations of $u(x)$ which also leave the spectrum invariant. The simplest of these is the Korteweg-de Vries,
\begin{equation}
\label{kdvequation}
4\partial_{t_3} u = 6u \partial_x u + \partial_x^3 u
\end{equation}
and was discovered by Boussinesq already in 1877. It is quite easy to formally check, for example using the so-called Lax equation
\begin{equation}
\partial_{t_3} L = \left[ \partial_x^3 + \frac{3}{2} u(x) \partial_x + \frac{3}{4} u^\prime(x), L \right],
\end{equation}
that the spectrum of \eqref{opschroedinger} is independent of the KdV time $t_3$. Unlike the rigid translation \eqref{eqtranslation}, the KdV equation \eqref{kdvequation} is a non-linear partial differential equation for $u(x)$ and leads to rather non-trivial deformation of the original Schr\"odinger potential $u(x)$. In fact, more is true. There exist infinitely many such continuous deformations and these are organized into a hierarchy of commuting flows analogous to \eqref{eqtranslation} and \eqref{kdvequation}. This family of flows is called the \emph{KdV hierarchy}.

There is a nice reformulation of the KdV hierarchy as a Hamiltonian system. We first consider an infinite dimensional space of all Schr\"odinger potentials. On this space we can define a Poisson bracket (Gelfand-Dickey bracket)
\begin{equation}
\label{gelfanddickeypoisson}
\left\{u(x),u(y)\right\} = -\delta^{\prime\prime\prime}(x-y)-4u(x)\delta^\prime(x-y)-2u^\prime(x)\delta(x-y).
\end{equation}
Although it might look complicated, it is natural from the point of view of classical Miura transformation. Furthermore, specializing to potentials living on a circle, the Poisson bracket \eqref{gelfanddickeypoisson} in terms of Fourier modes is simply the Virasoro algebra. Therefore the bracket \eqref{gelfanddickeypoisson} is the classical analogue of the quantum OPE \eqref{virasoroope}.

The equations of KdV hierarchy are generated by a collection of Poisson commuting Hamiltonians $I_{2j+1}$,
\begin{equation}
\partial_{t_{2j+1}} u = \left\{ I_{2j+1}, u \right\}
\end{equation}
with
\begin{equation}
\label{classicalkdvhamiltonians}
I_1 = \int u(x) dx, \qquad I_3 = \int u(x)^2 dx, \ldots.
\end{equation}
Since the flows of KdV hierarchy are commuting Hamiltonian flows, the quantities \eqref{classicalkdvhamiltonians} are also automatically conserved, i.e. the Hamiltonians such as \eqref{classicalkdvhamiltonians} are determined in terms of the spectral data associated to the Schr\"odinger operator \eqref{opschroedinger}.

The preceding discussion of KdV hierarchy can be generalized to so-called $N$-KdV hierarchy which is the classical analogue of quantum $\mathcal{W}_N$ algebras. One simply replaces the second order differential operator by an operator of $N$-th order. We still have infinitely many commuting flows preserving the spectrum of these higher order differential operators and the corresponding classical Hamiltonian system, but now the Hamiltonians and KdV times are not labeled by odd integers but by all integers and we do not have a single potential $u(x)$ but actually a $(N-1)$-tuple of these. The corresponding $\mathcal{W}_\infty$ version leads analogously to Kadomtsev–Petviashvili hierarchy of integrable partial differential equations.

\subsection{Quantum KdV charges}
Returning back to quantum field theory, a natural question to ask is what is the fate of the classical integrable structure of the KdV equation in the quantum setting. It turns out that the classical Hamiltonians \eqref{classicalkdvhamiltonians} have unique quantum analogues if we assume that they should be the zero Fourier modes of local quantities and commute with one another. For the first three of these we have
\begin{align}
I_1 & = \int T(x) dx = L_0 - \frac{c}{24} \\
I_3 & = \int (TT)(x) dx = L_0^2 + 2 \sum_{m>0} L_{-m} L_m - \frac{c+2}{12} L_0 + \frac{c(5c+22)}{2880} \\
I_5 & = \int \left[ (T(TT))(x) -\frac{c+2}{12} (\partial T \partial T)(x) \right] dx
\end{align}
The first two quantities are the only two non-trivial integrals of dimension $2$ and $4$ local fields that we can construct out of Virasoro stress-energy tensor $T(x)$. Starting from $I_5$, we have more independent local quantities, but there is always a unique combination that commutes with $I_3$ (up to an overall normalization) \cite{Bazhanov:1994ft}. Proceeding similarly, we find an infinite number of mutually commuting conserved quantities $I_{2j-1}$. In the classical limit $c \to \infty$ these quantities reduce to the classical KdV Hamiltonians \eqref{classicalkdvhamiltonians}.

The question is what is the spectrum of $\left\{I_{2j-1}\right\}$ in Virasoro representations. Since $I_1$ is up to a constant the Virasoro $L_0$, the problem of diagonalization of $\left\{I_{2j-1}\right\}$ reduces to a diagonalization of finite dimensional matrices level by level. But already at moderate levels it is quite hard to perform the diagonalization explicitly.

A suprising description of the spectrum of quantum KdV Hamiltonians was found by Bazhanov, Lukyanov and Zamolodchikov in \cite{Bazhanov:2003ni} in the context of ODE/IM correspondence \cite{Dorey:2007zx}. From the ODE/IM correspondence it was known that the primaries (highest weight states) of the Virasoro representations correspond to differential operators of the form
\begin{equation}
\label{odeimschroedinger}
-\partial_z^2 + \frac{\ell(\ell+1)}{z^2} + \frac{1}{z} + \lambda z^{\kappa}
\end{equation}
where $\kappa$ is related to the central charge $c$,
\begin{equation}
c = -\frac{(2\kappa+1)(3\kappa+4)}{\kappa+2}
\end{equation}
and $\ell$ to the conformal dimension $\Delta$ of the primary. The ODE/IM correspondence relates the data associated to the differential equation such as the Stokes data at infinity or the radial connection coefficients to spectra of local and non-local conserved quantities on the CFT side. In particular, using the WKB calculations applied to \eqref{odeimschroedinger} we can calculate the eigenvalues of $I_{2j-1}$ Hamiltonians of the primaries \cite{Dorey:2019ngq}.

In \cite{Bazhanov:2003ni} the authors generalized this identification also to all the Virasoro descendants. They allowed for additional singularities in the Schr\"odinger potential \eqref{odeimschroedinger} of the form
\begin{equation}
\sum_{j=1}^M \left( \frac{2}{(z-z_j)^2} + \frac{\gamma_j z_j}{z(z-z_j)} \right).
\end{equation}
In general, such a modification of the potential introduces a non-trivial monodromy of the wave functions around the singular points $z_j$. Requiring that the monodromy around these points is trivial for all values of $\lambda$ is a quantization condition that determines the parameters $z_j$ and $\gamma_j$. In particular the positions of the double poles $z_j$ have to satisfy a system of algebraic equations
\begin{multline}
{\small
\frac{\kappa(\kappa+2)^2}{4 z_j^3} - \frac{\kappa+1}{z_j^2} -\frac{(\kappa+2)\ell(\ell+1)}{z_j^3} = }\\
{\small = \sum_{k\neq j} \left( \frac{2\kappa}{z_j (z_j-z_k)^2} + \frac{\kappa}{z_k (z_j-z_k)^2} + \frac{\kappa^2}{z_j z_k (z_j-z_k)} - \frac{\kappa(\kappa+1)}{z_j^2 z_k} + \frac{4}{(z_j-z_k)^3} \right)}
\end{multline}
which we can think of as being Bethe equations for quantum KdV system. It is a BAE system of Calogero type (i.e. Gaudin type with degree $3$ interaction). This system of algebraic equations has finitely many solutions corresponding to simultaneous eigenstates of $\left\{I_{2j-1}\right\}$ at Virasoro level $M$ \cite{Conti:2020zft}. Given a solution of Bethe equations, the eigenvalues of all $\left\{ I_{2j-1} \right\}$ are given simply as symmetric polynomials of Bethe roots $z_j$.

Although this description of spectrum of quantum KdV Hamiltonians is quite concrete, it has some drawbacks. It is not completely straightforward to write explicitly such system for higher rank than $N=2$ (see \cite{Masoero:2019wqf} for $N=3$). Also, the discrete symmetries of $\mathcal{W}_\infty$ such as the Feigin-Frenkel duality and triality are not manifest. Finally, it is not clear how are the quantum KdV conserved quantities related to Yangian ones.

In the following we will use the algebraic Bethe ansatz to construct a one-parametric family of integrable structures which interpolate between the Yangian and quantum KdV integrable structures. Furthermore, we will write a corresponding system of Bethe ansatz equations which manifest all the discrete symmetries of $\mathcal{W}_\infty$ and which has a uniform form for any rank $N$ (and in fact even for all other truncations of $\mathcal{W}_{1+\infty}$).

\section{Miura transformation and instanton R-matrix}

\begin{figure}
\sidecaption
\begin{tikzpicture}[auto, node distance=2cm,>=latex']
\node [draw, rectangle, minimum height=3em, minimum width=6em] at (-2,1) (wsym) {$\mathcal{W}$-symmetry};
\node [draw, rectangle, minimum height=3em, minimum width=6em] at (2,1) (yangian) {Affine Yangian};
\node [draw, rectangle, minimum height=3em, minimum width=6em] at (0,-2) (miura) {Miura transformation};
\draw [latex-latex,left] (miura) -- node {free fields} (wsym);
\draw [latex-latex,right] (miura) -- node {$\mathcal{R}$-matrix} (yangian);
\end{tikzpicture}
\caption{Miura operator acting as a bridge between Yangian and $\mathcal{W}$-algebra description}
\label{miurafigure}
\end{figure}
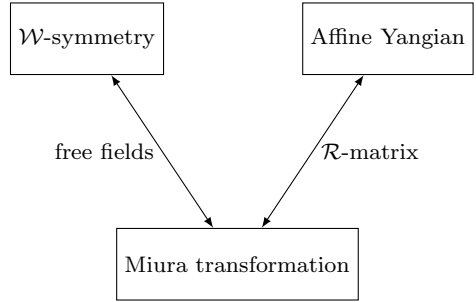

In this section we will return to Miura transformation \eqref{miura} and explain how it acts as a conceptual bridge between the Yangian and OPE description of the algebra. In fact, the Miura operator contains a large amount of algebraic information about the algebra. Apart from the free field representation that we discussed in Section \ref{secwalgebras}, it also lets us define the so-called instanton $\mathcal{R}$-matrix which is a central object that allows us to study the algebra from the point of view of quantum integrability.

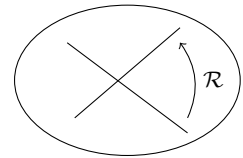
\begin{figure}
\sidecaption
\begin{tikzpicture}
\draw (0,0) ellipse (1.5 and 1);
\draw (-0.7,-0.5) -- (0.7,0.7);
\draw (-0.8,0.5) -- (0.8,-0.7);
\draw[->,bend right=30] (0.8,-0.5) to node[right] {$\mathcal{R}$} (0.7,0.5);
\end{tikzpicture}
\caption{Maulik-Okounkov instanton $\mathcal{R}$-matrix acting as an intertwiner between two free field representations of Virasoro algebra.}
\label{figinstRmatrix}
\end{figure}

Consider the rank $N=2$ case of the Miura operator \eqref{miura} acting on the Fock space $\mathcal{F}^{\otimes 2}$ of two commuting free bosons $J_1(z)$ and $J_2(z)$. Since the derivatives are acting to the right, the definition \eqref{miura} is asymmetric with respect to exchange of $J_1(z)$ and $J_2(z)$, i.e. we find two different embeddings of $\widehat{\mathfrak{u}}(1) \times \mathtt{Vir}$ in the Fock space $\mathcal{F}^{\otimes 2}$. But since the corresponding $N=2$ truncations of $\mathcal{W}_{1+\infty}$ are the same, there should exist an operator $\mathcal{R}:\mathcal{F}^{\otimes 2} \to \mathcal{F}^{\otimes 2}$ intertwining between these two embeddings,
\begin{equation}
(\alpha_0 \partial + J_1(z)) (\alpha_0 \partial + J_2(z)) = \mathcal{R}^{-1} (\alpha_0 \partial + J_2(z)) (\alpha_0 \partial + J_1(z)) \mathcal{R}.
\end{equation}
This is illustrated in Figure \ref{figinstRmatrix}. The object $\mathcal{R}$, sometimes called instanton $\mathcal{R}$-matrix, was introduced by Maulik and Okounkov in \cite{Maulik:2012wi} and later studied in \cite{Prochazka:2019dvu,Smirnov:2013hh,Zhu:2015nha}. It is closely related to Zamolodchikov's Liouville reflection operator that plays an important role in the study of quantum Liouville theory \cite{Zamolodchikov:2007zz}.

If we consider the situation with three free bosons, there are two ways of reordering $321 \to 123$. We can can obtain these by composing the three elementary $\mathcal{R}$-matrices and this leads to Yang-Baxter equation
\begin{equation}
\mathcal{R}_{12}(u_1-u_2) \mathcal{R}_{13}(u_1-u_3) \mathcal{R}_{23}(u_2-u_3) = \mathcal{R}_{23}(u_2-u_3) \mathcal{R}_{13}(u_1-u_3) \mathcal{R}_{12}(u_1-u_2)
\end{equation}
Here $\mathcal{R}_{jk}(u_j-u_k)$ is the elementary $\mathcal{R}$-matrix acting on $j$-th and $k$-th Fock space. We also wrote explicitly the \emph{spectral parameter} which in our case is simply the zero Fourier mode of $J_j(z)$, i.e. the global $\mathfrak{u}(1)$ charge (we can treat these as a spectral parameters since they are central).

Having found an $\mathcal{R}$-matrix satisfying the Yang-Baxter equation, we can immediatelly apply the algorithmic procedure of algebraic Bethe ansatz (quantum inverse scattering) to construct the Hamiltonians acting on the corresponding spin chain as well as to solve the resulting integrable system \cite{Faddeev:1996iy,Nepomechie:1998jf,Slavnov:2018kfx}. This is entirely analogous to the usual situation of say Heisenberg $su(2)$ XXX spin chain, except for the crucial difference that now the Hilbert spaces associated to every site of the spin chain are infinite dimensional (being the free boson Fock spaces). If we consider a spin chain of length $N$, the corresponding symmetry algebra will be the algebra $\widehat{\mathfrak{u}}(1) \times \mathcal{W}_N$, i.e. the usual Yangian level is now related to the rank (highest spin of the generators) of the resulting symmetry algebra.

In order to construct the Hamiltonians acting on our spin chain, we first introduce an additional \emph{probe} (or auxilliary) Fock space $\mathcal{F}_A$ and use the instanton $\mathcal{R}$-matrix to couple it to the Fock space $\mathcal{F}_Q \equiv \mathcal{F}_1 \otimes \ldots \mathcal{F}_N$ of $N$ free bosons. Using $\mathcal{R}$-matrix to couple the probe guarantees the consistency of the coupling with the symmetry algebra. The space $\mathcal{F}_Q$ is usually called the \emph{quantum space} in the integrable model literature. This coupling is captured by so-called \emph{monodromy} matrix $\mathcal{T}_{AQ}: \mathcal{F}_A \otimes \mathcal{F}_Q \to \mathcal{F}_A \otimes \mathcal{F}_Q$ defined as
\begin{equation}
\label{monodromymatrix}
\mathcal{T}_{AQ}(u_A) = \mathcal{R}_{A1}(u_A-u_1) \mathcal{R}_{A2}(u_A-u_2) \cdots \mathcal{R}_{AN}(u_A-u_N).
\end{equation}
The nice feature of $\mathcal{R}$-matrix satisfying Yang-Baxter equation is the reproducing property, the fact that its fused products such as \eqref{monodromymatrix} still satisfy a Yang-Baxter equation (i.e. we can think of the monodromy matrix as being the $\mathcal{R}$-matrix in more complicated representation). In order to see see this we introduce another probe, an auxiliary Fock space $\mathcal{F}_B$ and we define the associated monodromy matrix as in \eqref{monodromymatrix} but with probe label $A$ replaced by label $B$. Two such monodromy matrices satisfy the Yang-Baxter equation in the form
\begin{equation}
\label{monodromyybe}
\mathcal{R}_{AB}(u_A-u_B) \mathcal{T}_{AQ}(u_A) \mathcal{T}_{BQ}(u_B) = \mathcal{T}_{BQ}(u_B) \mathcal{T}_{AQ}(u_A) \mathcal{R}_{AB}(u_A-u_B).
\end{equation}
Here we are suppressing from the notation the spectral parameters associated to the quantum space (the zero modes of $J_1(z),\ldots,J_N(z)$). Taking specific matrix elements of the monodromy matrix \eqref{monodromymatrix} in the probe space leads to representation of Yangian generators such as $\psi(u), e(u)$ or $f(u)$ acting on the quantum space \cite{Prochazka:2019dvu}. On the other hand, matrix elements of \eqref{monodromyybe} in both of the probe Fock spaces express the quadratic Yangian relations.

The usual next step for the finite spin chains would be to take the trace of the monodromy matrix \eqref{monodromymatrix} over the probe Hilbert space to eliminate the probe and get an operator acting purely on the quantum space. The resulting object is called the transfer matrix $\mathcal{H}(u)$ and the trace of \eqref{monodromyybe} over both auxiliary spaces would show that the resulting quantities commute for any values of the spectral parameter,
\begin{equation}
\label{commutinghamiltonians}
\left[ \mathcal{H}(u), \mathcal{H}(v) \right] = 0.
\end{equation}
The Taylor coeffcients of $\mathcal{H}(u)$ would then give us commuting Hamiltonians acting on the spin chain.

\begin{figure}[t]
\sidecaption
\includegraphics[scale=0.35]{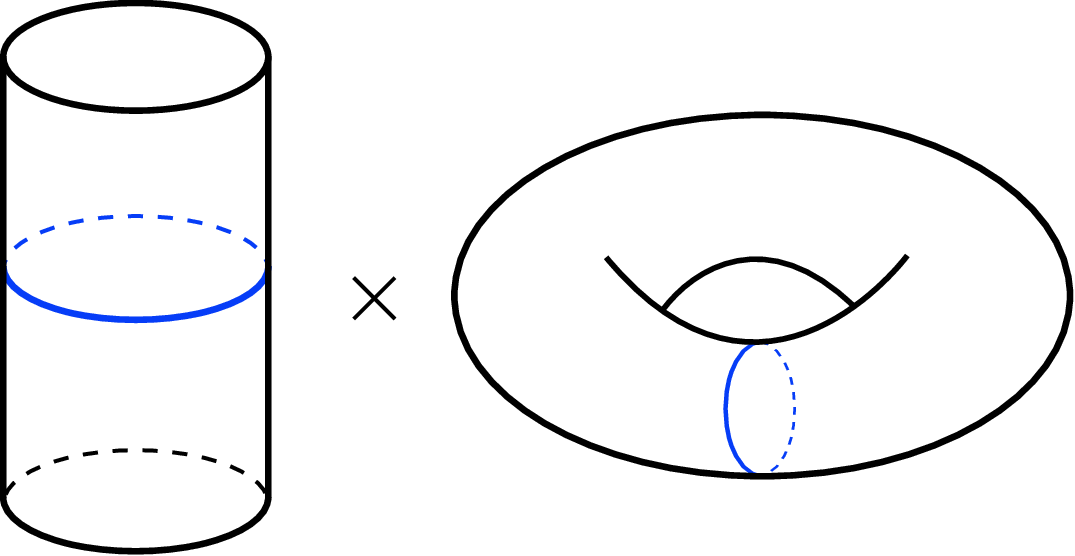}
\caption{The $\mathcal{R}$-matrix couples the auxiliary worldsheet to the cylinder where the Hamiltonians act (blue circles). The trace over the auxiliary space effectively puts the probe boson on a complex torus.}
\label{probetorus}
\end{figure}

In our situation, the auxiliary space associated to the probe is the bosonic Fock space so it is infinite dimensional. Therefore the trace of the monodromy matrix does not converge and needs to be regularized. A natural candidate for the regularized trace is
\begin{equation}
\label{ilwhamiltonians}
\mathcal{H}_q(u) \equiv \Tr_A q^{L_0^{(A)}} \mathcal{T}_{AQ}(u).
\end{equation}
In other words, we sum over all the Young diagrams representing the states in the probe Fock space and weight each contribution by $q$ to the power of the number of the boxes of the corresponding Young diagram. This is illustrated in Figure \ref{probetorus}. We can think of monodromy matrix to be a consistent coupling of the probe space to the cylinder on which we are studying the Hamiltonians. Regularization of the trace effectively puts the probe boson on a torus and the auxiliary parameter $q$ parametrizes the complex structure of the auxiliary torus. Importantly, the regularization does not spoil the crucial commutativity \eqref{commutinghamiltonians} and therefore we end up with a one-parametric family (parametrized by $q$) of integrable structures. It remains to identify the corresponding integrable systems.

By studying the large $u$ expansion of the Hamiltonians \eqref{ilwhamiltonians}, we find that apart from $L_0$, the first non-trivial Hamiltonian is a certain non-local deformation of \eqref{psi3yangian},
\begin{equation}
(W_3 + \ldots)_0 + \sigma_3 \sum_{m>0} m \frac{1+q^m}{1-q^m} J_{-m} J_m.
\end{equation}
This Hamiltonian has been identified in \cite{Litvinov:2013zda} with one of the Hamiltonians of the quantum intermediate long wave (ILW) hierarchy \cite{Buryak:2018ro}. It interpolates between the Yangian (or Benjamin-Ono) Hamiltonians to which it reduces in the limit $q \to 0$ (as well as $q \to \infty$) and the local quantum KdV or KP Hamiltonians in the $q \to 1$ limit. The limit $q \to 0$ is a low temperature limit from the probe point of view and the trace in \eqref{ilwhamiltonians} effectively reduces to a projection to the probe ground state. It is known that in this limit the matrix elements of the monodromy matrix give the Yangian quantities \cite{Prochazka:2019dvu}. On the other hand, the limit $q \to 1$ is the high temperature limit where all the states in the auxiliary Fock space contribute and we see that we get back the divergence that was previously regularized by introducing $q$. Nevertheless, by suitable rescaling one can systematically extract the finite local Hamiltonians \cite{Prochazka:2023zdb}.

Remarkably, for any value of $q$ there exists a system of Bethe ansatz equations that diagonalizes the ILW Hamiltonians \eqref{ilwhamiltonians}. The equations are \cite{Bonelli:2014iza,Kozlowski:2016too,Litvinov:2013zda,Nekrasov:2009rc}
\begin{equation}
\label{ilwbae}
1 = q \prod_{l=1}^N \frac{u_j+a_l-\epsilon_3}{u_j+a_l} \prod_{k \neq j} \frac{(u_j-u_k+\epsilon_1)(u_j-u_k+\epsilon_2)(u_j-u_k+\epsilon_3)}{(u_j-u_k-\epsilon_1)(u_j-u_k-\epsilon_2)(u_j-u_k-\epsilon_3)}
\end{equation}
Notice that they are strikingly similar to the Bethe equations of the simplest Heisenberg $SU(2)$ XXX spin chain. The deformation parameter $q$ has the role of the twist. The first product on the right-hand side represents the interaction of $j$-th Bethe root $u_j$ with external fields. In our case this interaction is controlled by Coulomb parameters $a_l$ which parametrize the highest weights of the primary. The second product is the interaction of $j$-th Bethe root with all other roots and it is determined by the structure function \eqref{structurefunction}. This function is the main difference between ILW BAE for $\mathcal{W}_{1+\infty}$ and the $\mathfrak{su}(2)$ Heisenberg spin chain: \eqref{structurefunction} is a degree $3$ function with three zeros and three poles while the usual interaction term in Heisenberg chain is of degree one.

Even though the equations \eqref{ilwbae} look quite simple, they have extremely rich structure of solutions -- in fact they need to capture the representation theory of Virasoro of $\mathcal{W}_N$ algebras such as their minimal models, structure of the null states for the degenerate primaries etc. \cite{Prochazka:2023zdb}.

In general the equations \eqref{ilwbae} are difficult to solve even numerically. But in the low probe temperature limit $q \to 0$ or $q \to \infty$ the study of solutions reduces to study of zeros of the numerators or denominators in \eqref{ilwbae}. Since these are factorized into product of linear factors, this in turn leads to equations such as $u_j = u_k \pm \epsilon$ and we immediately recognize the process of growing of the plane partition. The Bethe roots as $q \to 0$ are associated to individual boxes of the 3d Young diagrams.

The parameter $q$ has several natural roles:
\begin{enumerate}
\item from the spin chain point of view it is the twist parameter, controlling the phase change of the wave function as we close the chain
\item from the algebraic Bethe ansatz point of view it encodes the conformal structure of the auxiliary torus so the temperature of the probe particle
\item from the point of view of ILW Hamiltonians, it controls their non-locality and allows us in particular to interpolate between the local and Yangian Hamiltonians
\item finally when it comes to actual numerical solution of the equations, $q$ is a very convenient homotopy parameter -- it connects the Yangian $q \to 0$ regime where the equations can be solved exactly to the regime of finite $q$
\end{enumerate}

Given a solution of ILW Bethe ansatz equations the joint eigenvalues of $\mathcal{H}_q(u)$ can be written as
\begin{equation}
\frac{\mathcal{H}_q(u)}{\mathcal{H}_{q=0}(u)} \to \frac{1}{\sum_\lambda q^{|\lambda|}} \sum_\lambda q^{|\lambda|} \prod_{\Box\in\lambda} \psi_\Lambda(u-\epsilon_\Box+\epsilon_3)
\end{equation}
where
\begin{equation}
\psi_\Lambda(u) = A(u) \prod_j \varphi(u-x_j).
\end{equation}
where $A(u)$ encodes the highest weights. An analogous formula has been derived by Feigin, Jimbo, Miwa and Mukhin in the context of $q$-deformed versions of $\mathcal{W}_{1+\infty}$ \cite{Feigin:2016pld,Feigin:2016uxd} and further studied in \cite{Feigin:2017gcv,Prochazka:2023zdb}.

Even though the system of equations \eqref{ilwbae} provides a nice answer to the problem of finding spectrum of the integrable system associated to $\mathcal{W}_{1+\infty}$, there are still many questions to be answered, for example the relation to the BLZ like systems of Bethe equations, relation to the associated quantum elliptic Calogero models \cite{Nekrasov:2009rc}, the relation between low and high energy properties and modularity etc. Even the characterization of the singular behaviour of Bethe roots as $q \to 1$ in terms of excitations of the free boson is very rich and not quite trivial \cite{Prochazka:2024abc}!

\begin{acknowledgement}
I would like to thank to organizers of all workshops and seminars where this work was presented. Special thanks to Christopher Beem, Giulio Bonelli, Ilka Brunner, Dylan William Butson, Stefan Fredenhagen, Alba Grassi, Harald Grosse, Saebyeok Jeong, Branislav Jur\v{c}o, Taro Kimura, Shota Komatsu, Mat\v{e}j Kudrna, Oleg Lisovyy, Alexey Litvinov, Yutaka Matsuo, Davide Masoero, Go Noshita, Valentina Petkova, Miroslav Rap\v{c}\'{a}k, Martin Schnabl, Alessandro Tanzini, Miroslav Ve\v{l}k, Akimi Watanabe, Jingxiang Wu, Ida Zadeh for useful discussions. The research was supported by the Grant Agency of the Czech Republic under the grant EXPRO 20-25775X.
\end{acknowledgement}
\end{document}